\documentclass[aps,english,10pt,superscriptaddress,twocolumn]{revtex4}  
\usepackage{amsmath}
\usepackage{amsthm}
\usepackage{amssymb}
\usepackage{amsfonts}
\usepackage{bbm}
\usepackage{epsfig}
\usepackage{times}
\usepackage{babel}
\usepackage{color}
\usepackage{hyperref}
\usepackage{framed}
\usepackage{changes}
\usepackage{float}

\def\tr{{\rm tr}}

\begin{document}

\title{Out of Time Ordered Correlators and Entanglement Growth in the Random Field XX Spin Chain}
\author{Jonathon Riddell}
\email{riddeljp@mcmaster.ca}
\affiliation{Department of Physics \& Astronomy, McMaster University
1280 Main St.  W., Hamilton ON L8S 4M1, Canada.}
\author{Erik S. S{\o}rensen }
\email{sorensen@mcmaster.ca}
\affiliation{Department of Physics \& Astronomy, McMaster University
		1280 Main St.  W., Hamilton ON L8S 4M1, Canada.}
	
\date{\today}

\begin{abstract}
We study out of time order correlations, $C(x,t)$ and entanglement growth in
the random field XX model with open boundary conditions using the exact
Jordan-Wigner transformation to a fermionic Hamiltonian. For any non-zero
strength of the random field this model describes an Anderson
insulator.  Two scenarios are considered: A global quench with the
initial state corresponding to a product state of the N\'eel form, and
the behaviour in a typical thermal state at $\beta=1$. As a result of
the presence of disorder the information spreading as described by the
out of time correlations stops beyond a typical length scale,
  $\xi_{OTOC}$. For $|x|<\xi_{OTOC}$ information spreading occurs at the maximal velocity $v_{max}=J$ and 
we confirm predictions for the early time behaviour of $C(x,t)\sim t^{2|x|}$. 
For the case of the quench starting from the N\'eel product state
we also study the growth of the bipartite entanglement, 
focusing on the late and infinite time behaviour.  The approach to a bounded entanglement is observed to be slow for
the disorder strengths we study.  
\end{abstract}

\maketitle

\section{Introduction}
A recent conjecture~\cite{Maldacena2016a} establishing a bound for the rate of growth of chaos in quantum systems
have spurred interest in the study correlators of the form~\cite{Larkin1969}:
\begin{equation}
        C(x,t)=\langle [W(x,t),V(0)]^\dagger[W(x,t),V(0)]\rangle,
\label{eq:otocgeneral}
\end{equation}
where $W$ and $V$ are local non-overlapping operators separated by a displacement $x$, $[W(x,0),V(0)]=0$, and $\langle\cdot\rangle$ is a thermal average.
If $W,V$ are both hermitian and unitary it follows that
\[
	C(x,t) = 2(1-\Re[F(x,t)])
\]   with $F(x,t)=\langle W(x,t)V(0)W(x,t)V(0)\rangle$
and $F$ is therefore referred to as an out-of-time-ordered correlator (OTOC).
While $W$ and $V$ commute at $t=0$ this may no longer be the case at a later time giving rise to the notion of a growing ``operator radius"~\cite{Swingle2017}
defined as the distance, $R_W(t)$, where $F(x,t)$ significantly deviates from 1 for all $|x|<R_W(t)$.
$C(x,t)$ can then be seen as a measure of the degree of non-commutativity of $W(x,t)$ and $V(0)$ for $t>0$ and if $C(x,t)$ remain large for an extended
period of time the system is said to be scrambled.

The time where $C(x,t)$ becomes ${\cal O}(1)$ defines a ``scrambling" time,
$t_*$, and for the early time approach to scrambling it is expected that for some models
$C(0,t)\sim e^{\lambda_L t}$ with the conjectured~\cite{Maldacena2016a} bound
$\lambda_L \le 2\pi k_BT/\hbar$. Systems that approach this bound are known as
fast
scramblers~\cite{Sekino2008,Sachdev1993,Sachdev2015,Roberts2015a,Fu2016,Maldacena2016b}.
This is in contrast to a range of models that do not exhibit this early time
exponential
growth~\cite{Dora2017,Huang2017,Swingle2017,Chen2017,Slagle2017,Fan2017,Deng2017}
and are therefore known as slow scramblers. In particular OTOCs in many-body
localized systems~\cite{Nandkishore2015,Alet2018} (MBL) have been
studied~\cite{Huang2017,Swingle2017,Chen2017,Slagle2017,Luitz2017,Fan2017,Xu2018a,Xu2018b,Sahu2018}
and early time power-law growth of $C(x,t)$ is
expected~\cite{Swingle2017,Huang2017,Chen2017,Fan2017} in such systems.
distinguishing them from Anderson localized (AL) models where $C(x,t)$ is
expected to be a constant~\cite{Fan2017}, at least for very strong disorder. 
The behaviour of the correlator $C(x,t)$ is therefore
capable of distinguishing different phases.
		
More generally, if the spatial dependence is taken into account, $C(x,t)$
exhibits the butterfly effect~\cite{Shenker2014a,Gu2017,Patel2017} with certain
models exhibiting the behaviour $C\sim e^{\lambda_L(t-x/v_B)}$. Here, $v_B$ is
the butterfly velocity that can be viewed as the velocity of information in
a strongly correlated systems.  
Perturbative weak coupling
calculations~\cite{Chowdhury2017,Patel2017} recover similar exponential
behaviour whereas random circuit
models~\cite{Nahum2018,Khemani2018,Rakovsky2018} show a diffusively spreading
$C\sim e^{-\lambda_L(x-v_Bt)^2/t}$ and for non-interacting translationally
invariant systems it can be shown that~\cite{Xu2018b,Xu2018a} $C\sim
e^{-\lambda_L(x-v_Bt)^{3/2}/t^{1/2}}$. A universal form has also been  
proposed~\cite{Xu2018b}.  
\begin{equation}
        C(x,t)\sim \exp\left(-\lambda_L\frac{(x-v_Bt)^{1+p}}{t^p}\right).
        \label{eq:uniC}
\end{equation}
It should be noted that these different forms are only expected to
be valid close to the ``wave-front'', where $x-v_Bt$ is small. 
We also note that, in general, $v_B$ can be different from
$v_E$~\cite{Liu2014}, the rate at which entanglement spreads, but for the
models we shall consider here $v_B=v_E$~\cite{Hosur2016}.

Recent studies~\cite{Hosur2016,Fan2017}, have also shown that $C(x,t)$ 
can be directly related to the second R\'enyi entropy
$S^{(2)}$ of an appropriately defined sub-system, and scrambling in
a quantum channel can be defined in terms of the tripartite information of a sub-system\cite{Hosur2016}.
The quasi-probability behind the OTOC~\cite{Halpern2017a,Halpern2017b,Alonso2018} has also been studied.

The closely related concept of the growth of entanglement after a quench has been intensely studied with
the observation of a logarithmic growth with time~\cite{BardarsonEnt,Serbyn2013,Deng2017} as one of the hallmark features of MBL.
In contrast, a thermal phase should exhibit linear growth of the entanglement and in the AL phase a bounding constant entanglement is expected~\cite{Abdul-Rahman2016,Fan2017}.

The relationship between scrambling, the OTOC and thermalization has also been
considered~\cite{VonKeyserlingk2018,Bohrdt2017,LewisSwan2018}.  Models which
can be mapped to a quasi-free fermionic model with de-localizing dynamics have
been studied showing that local 2-point correlation functions equilibrate to a
generalized Gibbs ensemble~\cite{Murthy2018, Gluza2018}.  An interesting
question is then, what signatures of generalized thermalization appear in an OTOC?

There are therefore many aspects that make the OTOC an object of considerable
current interest and exact numerical results are of significant interest in
particular in the presence of disorder. Previous
studies~\cite{Luitz2017,Fan2017} have in particular focused on MBL systems
where both disorder and interactions play an important role and severely limits
the sizes that can be reached in numerical calculations. If interactions are
neglected the Jordan-Wigner transformation can be used to study OTOCs. In the
absence of disorder such studies have been performed on the quantum Ising
chain~\cite{LinOTOCising}, quadratic fermions \cite{Byju2018} and hard-core boson models~\cite{Lin2018}. In ~\cite{LinOTOCising}
scrambling was observed at the critical point of the quantum Ising model in the OTOC for operators
non-local in the Jordan-Wigner fermions.

Here we turn the attention to the one-dimensional XX spin chain with a random field (RFXX), 
\begin{equation}
	\hat{H} = J \sum_{i=0}^{L-2} \left(S_{i}^{x}S_{i+1}^{x} + S_{i}^{y}S_{i+1}^{y}\right)  + \sum_{i=0}^{L-1} \lambda_i S_{i}^{z},
        \label{eq:Hxx}
\end{equation}
where $S_i^x,S_i^y$ and $S_i^z$ are the spin-$1/2$ operators at site $i$, $L$ is
the number of sites, $J$ is the interaction coefficient and the $\lambda_i$ are
the on site fields applied to the z-axis. The $\lambda_i$ are taken uniformly
from the interval $\lambda_i \in [-\lambda, \lambda]$ and we set $\hbar=1$.  We
shall refer to $\lambda$ as the disorder parameter and we shall mainly be
concerned with the weak disorder regime $\lambda<J$.  This model describes a
typical Anderson insulator and is in the AL phase for any non-zero $\lambda$.
This model is known to be dynamically localized~\cite{Hamza2012} in the sense that
it satisfies a zero-velocity Lieb-Robinson bound. Furthermore, entanglement is bounded 
at all times for this model~\cite{Abdul-Rahman2016}. However, relatively little is known
about the early-time behaviour in the model which is the focus of the present paper.
As we detail below, the Jordan-Wigner transformation is applicable to the random field XX spin chain
also in the presence of disorder and sizeable systems can be treated. To
simplify the calculation we exclusively consider open boundary conditions
(OBC). We focus on two different scenarios: A quench from a simple N\'eel like
product state with no entanglement of the following form
\begin{equation} \label{prodstate1}
		|\psi \rangle  = \prod_{l\in \mathbb{S}} \hat{S}_l^+ |\downarrow \rangle,
\end{equation}
where $\mathbb{S} = \{ l \in \mathbb{N}: l \mod 2 = 0 \}$. The second scenario corresponds to  a typical thermal state
\begin{equation} 
\rho = \frac{e^{-\beta \hat{H}}}{Z}.
\end{equation} 
with $\beta=1$ and $Z=\tr\ \exp(-\beta \hat{H})$. Expectation values for the two scenarios are then determined as:
\begin{equation}
  \langle O \rangle_{Neel}=\langle \psi |O|\psi\rangle,\ \ \mathrm{and}\ \  \langle O\rangle_{th} = \tr(\rho O).
\end{equation}

Our principal findings are the followings. The propagation of the OTOCs
essentially stops beyond a length $\xi_{OTOC}$ that depends on the strength of
the disorder, $\lambda$. For $|x|>\xi_{OTOC}$ $C(x,t)$ is essentially a
constant, $C(x,t)$ in agreement with previous studies~\cite{Fan2017} performed
at strong disorder and very small $\xi_{OTOC}$. However, for $|x|<\xi_{OTOC}$
the OTOC propagates information with the {\it maximal} group velocity $v_{max}
= J$ in the thermodynamic limit.  This is the case for both the product and
thermal state. For modest $\lambda$ $\xi_{OTOC}$ can be sizeable.  For
$|x|<\xi_{OTOC}$ the early-time regime of $C(x,t)$ is shown to behave as
$t^{2|x|}$ in accordance with a recent proposal~\cite{LinOTOCising}, even in
the presence of disorder, $\lambda\neq 0$. For $\lambda\neq 0$ the light-cone
therefore has the shape of a neck-tie with a v-shaped tip.  While the bipartite
entanglement in the RFXX model is bounded at all times~\cite{Abdul-Rahman2016}
we find that the approach to this bound at small $\lambda$ is rather slow.

The plan of the paper is as follows. In section~\ref{Secmodel} we outline some
technical aspects of applying the Jordan-Wigner transformation.  Section
\ref{Sec:OTOCs} presents our results for the OTOCs for the two different
scenarios detailed above and in section~\ref{Ent} discuss our results for the
evolution of the entanglement after a quench from the N\'eel product state.
Finally, in section~\ref{Sec:Loc} we attempt to extract a localization length
from the bipartite entanglement entropy.

\section{Jordan-Wigner Transformation}\label{Secmodel}
In order to study the model Eq.~(\ref{eq:Hxx}) we employ the
Jordan-Wigner transformation
\cite{Coleman}. Using $S_i^{\pm} = (S_i^x\pm i S_i^y)/2$,
\begin{eqnarray}
  S_i^+ &=& \prod_{j=1}^{i-1} \left(1-2 \hat f_j^\dagger \hat f_j\right)\hat f_i^\dagger,\quad S_i^- = \prod_{j=1}^{i-1} \left(1-2 \hat f_j^\dagger \hat f_j\right) \hat f_i,\nonumber\\
		S_i^z &=& \hat f_i^\dagger \hat f_i-\frac 1 2,
\end{eqnarray}
we recover a Hamiltonian, 
\begin{equation}
	\hat{H} = \frac{J}{2} \sum_{i=0}^{L-2}\left( \hat{f}_i^{\dagger}\hat{f}_{i+1} + \hat{f}_{i+1}^{\dagger}\hat{f}_i\right) 
        + \sum_{j=0}^{L-1} \lambda_j \left(\hat{f}_{j}^{\dagger}\hat{f}_{j}-\frac{1}{2}\right),
\end{equation}
which is a quasi-free fermionic Hamiltonian with anti-commutation relations  
$\{\hat{f}_k,\hat{f}_l \} =\{\hat{f}_k^{\dagger},\hat{f}_l^{\dagger}\}= 0$ and $\{\hat{f}_l^{\dagger},\hat{f}_k\} = \delta_{l,k}$. 
We adjust the spectrum to get rid of the constant term and write, 
\begin{equation}
		\hat{H} = \sum_{i,j}M_{i,j}\hat{f}_{i}^{\dagger}\hat{f}_{j} .
\end{equation}
Where $M$ is the effective Hamiltonian with entries $ M_{i,i} = \lambda_i$ and $ M_{i,j} =\frac{J}{2} $ if $|i-j|=1$. 
All other entries are zero. 
This model can be used to study differences between a thermal phase, with no disorder $\lambda=0$, and the localized phase with $\lambda\neq0$. 
When $\lambda = 0$ and we restrict ourselves to the case of $\langle \hat{N} \rangle = \sum_i^L \langle \hat{f}_i^\dagger \hat{f}_i \rangle = \frac{L}{2}$ a regime where 
the eigenstates of this model typically look locally identical to the Gibbs state [\cite{RiddellGETH}, \cite{Lai2015}]. 
However when $\lambda>0$ the eigenstates are localized and have exponentially decaying correlations characterized by some localization length \cite{Abdul-Rahman2016,RahmanXY,StolzALintro}.
	
Since $M$ is real symmetric, for a given field realization we can always diagonalize, $M=ADA^T $ where $AA^T = \mathbb{I}$ and $D$ is a diagonal matrix with entries $D_{k,k} = \epsilon_k$. Defining new fermionic operators,
\begin{equation}
	\hat{d}_k = \sum_j A_{j,k} \hat{f}_j.
\end{equation}
\begin{equation}
	\hat{d}_k^\dagger = \sum_j A_{j,k} \hat{f}_j^\dagger,
\end{equation}
we can then write the Hamiltonian as, 
\begin{equation}
	\hat{H} = \sum_{k}\epsilon_{k}\hat{d}_{k}^\dagger\hat{d}_{k},
\end{equation}
where the $\epsilon_k$ are the eigenmodes.  A simple reorganization and
applications of Wick's theorem when appropriate allows us to express out of
time ordered correlators in terms of two point correlations.  More details on
evaluating the time evolution of this model is presented in Appendix
\ref{TimeEvolve}. 
	
The problem of locality should be addressed. The Jordan-Wigner transformation
does not completely conserve locality, the $j$th pair of fermionic operators
are built from the $1, \dots j$ site spin operators, making it quasi-local.
However the $\hat{S}_i^z$ spin operators are mapped locally to fermions, so we
use these operators in the OTOC. Similarly, for the entanglement entropy we
consider subregions $A = \{1, \dots |A|\}$ which are blocks of spin sites
preserved by the transformation.  We have not considered OTOCs that are not
local in the fermion representation as was considered for the quantum Ising
model in Ref.~\cite{LinOTOCising}.

In the following we mainly focus on the disorder strength $\lambda = 0,0.3,0.8$
and we always fix $J=1$ and $\hbar=1$. We exclusively consider open boundary
conditions.  For the results presented in the following sections we typically
use a system size of $L=400$ and unless otherwise noted 1,000 disorder
realizations of the Hamiltonian are considered and averaged over. We use a
simple average to extract mean values over the disorder, leaving a study of the complete distribution
over the disorder for further study. When presenting
results for several time-slices of $C(x,t)$ each value of $C(x,t)$ is shifted
vertically by a value of $0.25t$ for visualization purposes.

\section{Out of time order correlations}
	\label{Sec:OTOCs}
In this section we investigate the out of time ordered correlations of the form,
\begin{equation}
\label{otoc}
  C(x,t) = \langle [\hat{\sigma}_i^z(t),\hat{\sigma}_j^z ]^\dagger  [\hat{\sigma}_i^z(t),\hat{\sigma}_j^z ] \rangle,
\end{equation}
where $x=i-j$ is understood to be the displacement between site $i$ and $j$.
Since $\hat{\sigma}_i^z$ is unitary we may write,
\begin{equation}
	C(x,t) = 2(1-\Re[F(x,t)]).
\end{equation}
We note that with this definition of $C(x,t)$ the maximum value it can reach is 2.
Here,
\begin{equation}
\label{fOTOC}
  F(x,t) = \langle \hat{\sigma}_i^z(t) \hat{\sigma}_j^z \hat{\sigma}_i^z(t) \hat{\sigma}_j^z  \rangle.
\end{equation}
We will fix the position of the time evolved operator as $i=\frac{L}{2}$.
Varying $j$ allows us to observe the operator radius spreading over the
lattice.  As described above, we consider two scenarios. A product state
generated by a set of creation operators where $\mathbb{S} = \{ l \in \mathbb{N}: l \mod 2 = 0 \}$. 
\begin{equation} \label{prodstate}
 |\psi \rangle  = \prod_{l\in \mathbb{S}} \hat{S}_l^+ |\downarrow \rangle = \prod_{l\in \mathbb{S}} \hat{f}_l^\dagger |0 \rangle,
\end{equation}
$|\downarrow \rangle$ and $|0\rangle$ are the all spin down and the vacuum
state respectively.  This state is a classical N\'eel state which has the
advantage of yielding essentially symmetric initial conditions for spins
surrounding the middle lattice point $i = \frac{L}{2}$ allowing us to restrict
our studies to one directional displacement on the lattice and having initial
fermions distributed evenly in real space.  For the second scenario of a
thermal state, we construct the Gibbs state with an inverse temperature $\beta
= 1$. More details on how these initial conditions are handled and how $C(x,t)$
is calculated can be found in Appendix~\ref{OTOCcalcs}. 

\begin{figure}[!ht]
\centering
  \includegraphics[width=\linewidth]{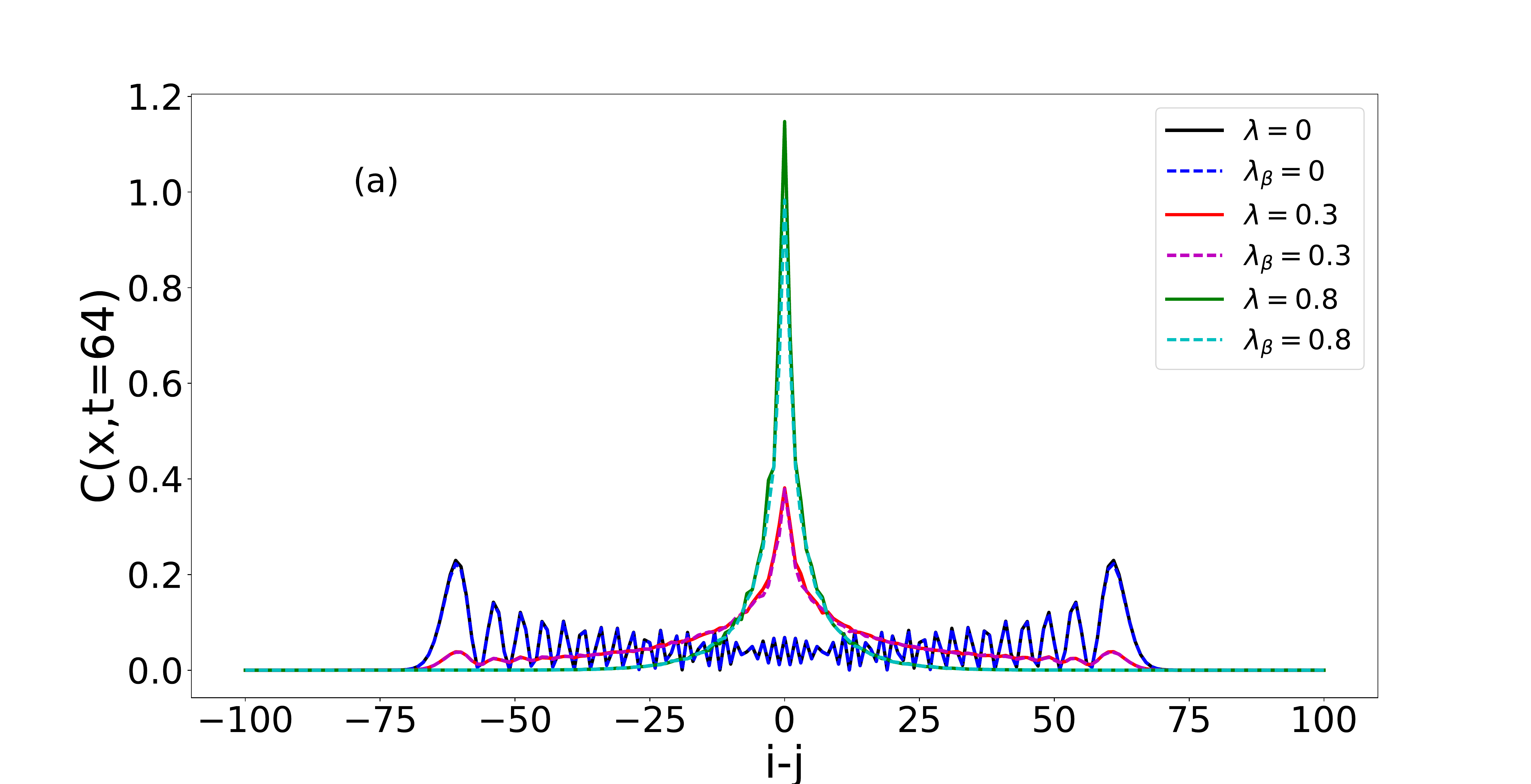}
  \includegraphics[width=\linewidth]{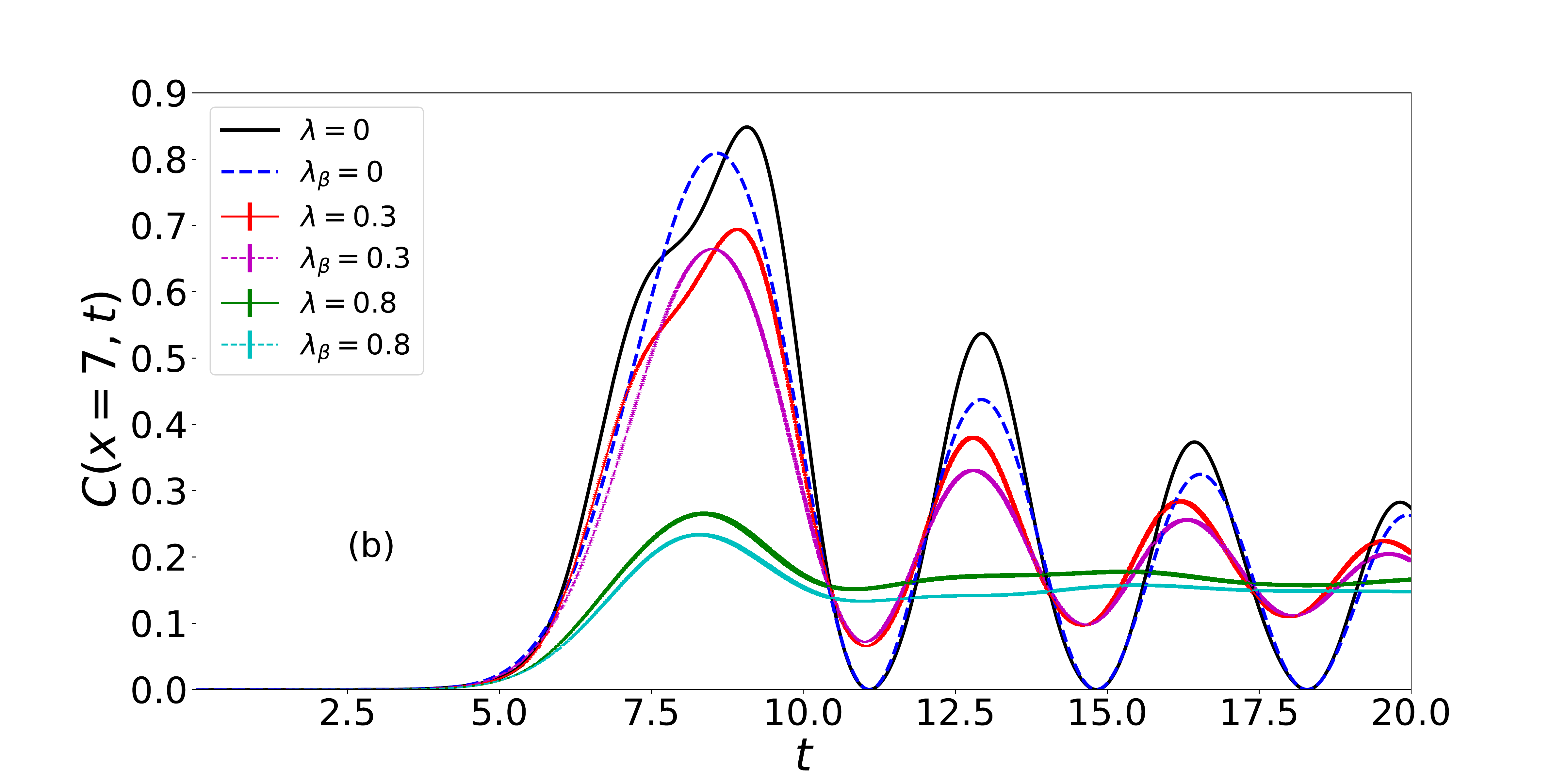}
  \includegraphics[width=\linewidth]{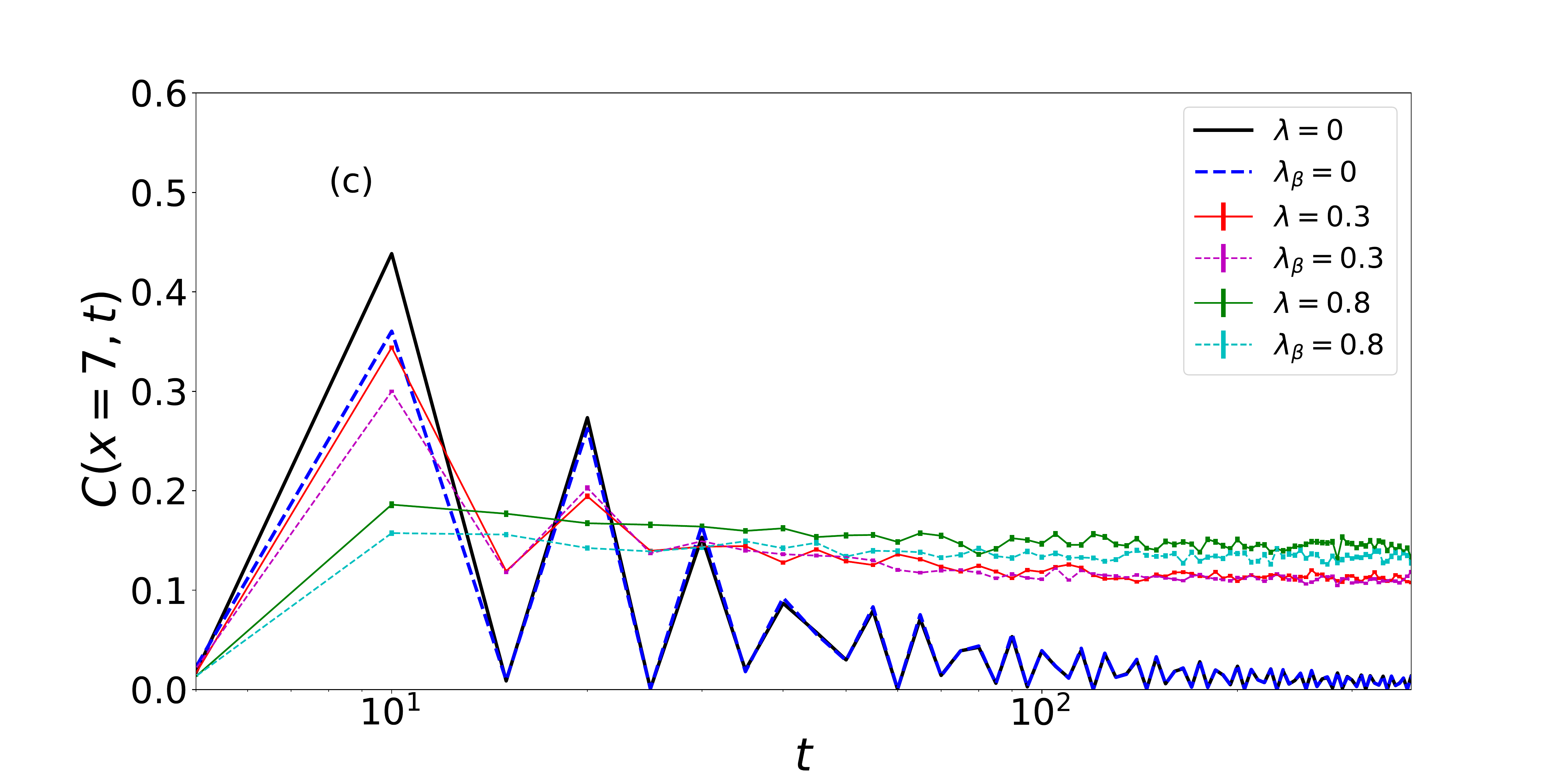}
  \caption{ Results for $C(x,t)$ for three different disorder strengths $\lambda=0,0.3,0.8$. 
  Comparing the product state and thermal state.
  The labelling $\lambda_\beta$ refers to $C(x,t)$ calculated in the thermal state with the specified disorder strength.
  Solid lines are results for the product state, dashed lines refer to the thermal state at $\beta=1$.
  (a) $C(x,t=64)$ versus $x$ for a fized $t=64$, shown as green line in Fig.~\ref{fig:OTOClam} and \ref{fig:thermOTOClam}. 
  (b) Early time behaviour of $C(x=7,t)$ at $x=7$, shown as the solid red line in Fig.~\ref{fig:OTOClam} and \ref{fig:thermOTOClam}.
  (c) Late time behaviour of $C(x=7,t)$.}
\label{fig:cx7tcompare}
\end{figure}
Before a more detailed discussion of our results for the two different
scenarios we discuss general features of the results for the OTOC and compare
the two scenarios in Fig.~\ref{fig:cx7tcompare} (solid lines represent results
for the product state, dashed lines for the thermal state).  Here, Fig.~\ref{fig:cx7tcompare}(a) show results  $C(x,t=64)$ at a fixed time $t=64$
versus $x$.  For both the thermal and product state the effects of the disorder
is immediately noticeable in the smoothening of $C(x,t)$ that is
characteristically oscillating with $x$ in the absence of disorder. For
$\lambda\neq 0$ $C(x,t)$ is sharply peaked around $x=0$ and a clear signature
of a wave-front where $C(x,t)$ first becomes non-zero is starting to disappear
for $\lambda=0.8$ for this time-slice. Fig.~\ref{fig:cx7tcompare}(b) show
results for $C(x=7,t)$ at a fixed separation $x=7$ versus time. Clear
differences between the results for the thermal state and the product state are
visible. Most notably, additional structure appear in the peaks of $C(x=7,t)$
for the product state while the thermal state yields a much smoother
oscillation.  The long-time behaviour of $C(x=7,t)$ is shown in
Fig.~\ref{fig:cx7tcompare}(c). While $C(x=7,t)$ clearly goes to zero for
$\lambda=0$ for both scenarios, indicating absence of scrambling, it appears
plausible that it attains a finite value in the long-time limit for
$\lambda=0.3,0.8$ for both scenarios. Since $C(x=7,t)$ does not saturate for
$x=7$ one could consider this weak (partial) scrambling for $\lambda=0.3,0.8$. We note
that there is a rather large variation in $C(x=x_0,t)$ with $x_0$ and as we
discuss below $C(|x|>\xi_{OTOC},t)$ is essentially zero for
{\it all} $t$ when $\lambda\neq 0$ indicating the absence of scrambling beyond this length scale.

We now turn to a more specific discussion of our results for the N\'eel product state and thermal state.

\subsection{Product States}
\begin{figure}[!ht]
\centering
  \includegraphics[width=\linewidth]{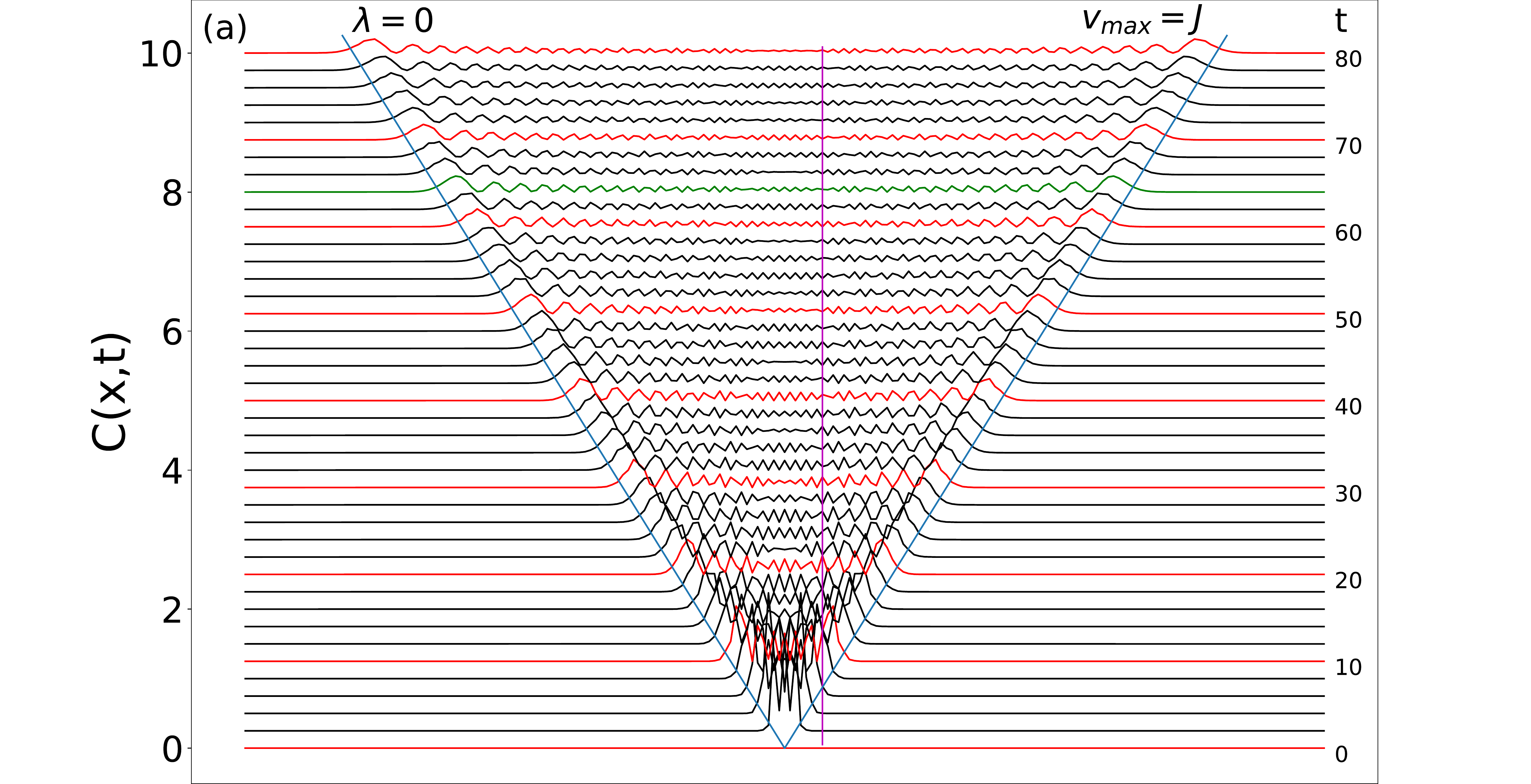}
  \includegraphics[width=\linewidth]{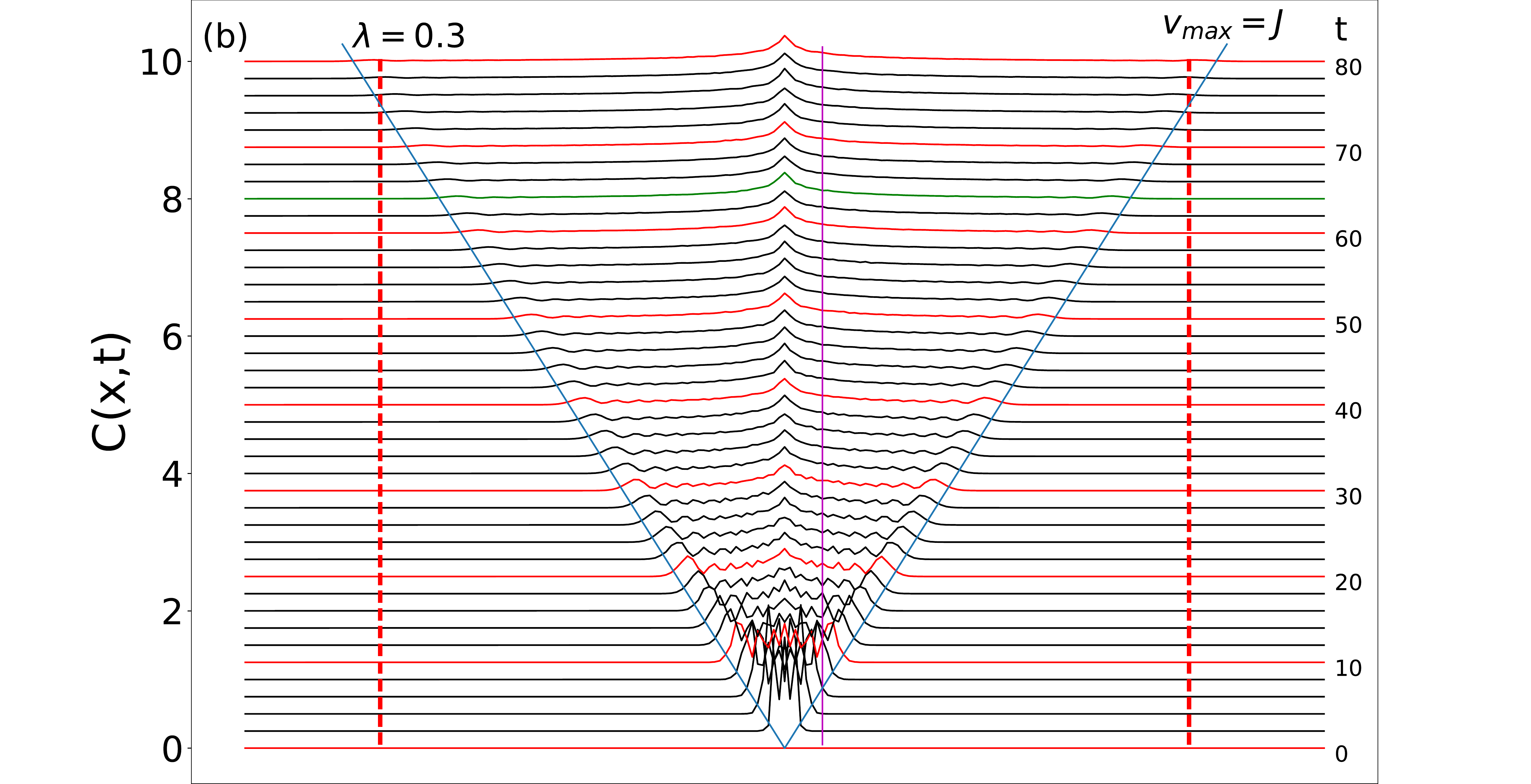}
  \includegraphics[width=\linewidth]{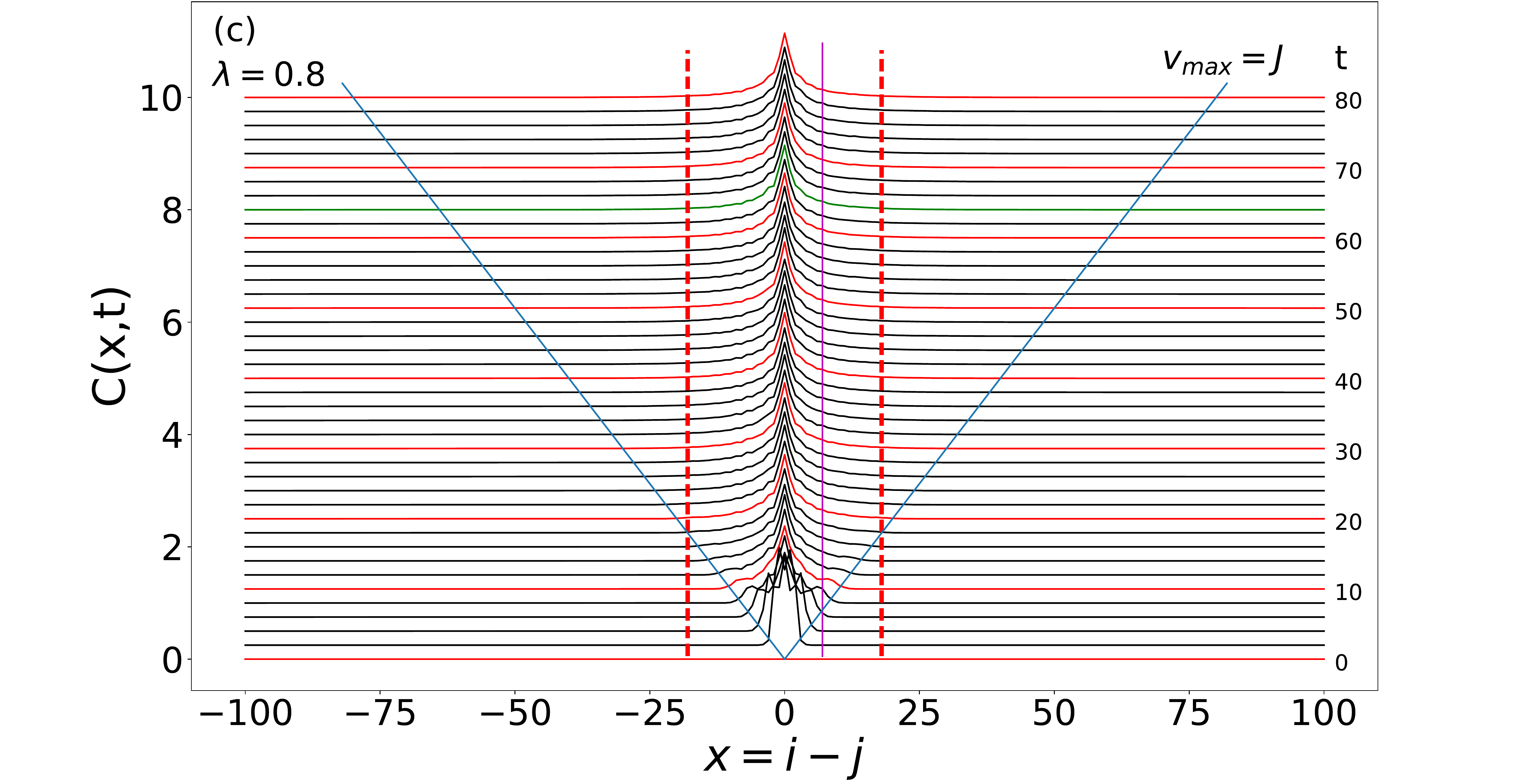}

\caption{Wave propagation plot of $C(x,t)$ for the XX spin model at disorder
        strength (a) $\lambda = 0$, (b) $\lambda=0.3$ and (c) $\lambda=0.8$. 
        For visualization, each value of $C(x,t)$ is
        shifted vertically by a value of $0.25t$ demonstrating
        the operators radius spreading. The $x$-axis is the displacement from
        the centre of the chain $i=\frac{L}{2}$. The two $y$-axis are
        the values $C(x,t)$ and the corresponding time. The maximal group
        velocity $v_{max} = J$ is also shown (solid blue line). 
        In panel (b) and (c)
        the vertical dashed red line indicates $\xi_{OTOC}$,  the $x$ value beyond which $C(x,t)< 10^{-3}$
        for any $x$. $\xi_{OTOC}= 18$ for $\lambda = 0.8$ and 75 for $\lambda =0.3$.
}\label{fig:OTOClam}
\end{figure}
In Fig.~\ref{fig:OTOClam} we show different time slices of $C(x,t)$
versus $x$. This shell like structure is expected and
parallels the results seen in Ref. \cite{LinOTOCising} for the quantum Ising
chain when constructing the OTOC with two operators which are local in the
fermionic representation. However, key differences emerge when disorder is introduced
by increasing $\lambda$. When $\lambda = 0$ we are in a
thermal phase and we observe operator spreading over the lattice in the sense that $C(x,t)$ eventually becomes becomes non-zero 
for any $x$ for large enough $t$. The operator
spreads over the lattice at the maximal group velocity $v_{max} = J$ as
expected. For an individual $x$ the $C(x,t)$ grows initially in time, peaks and
returns to zero with some rebounding with weaker peaks. (See Fig.~\ref{fig:cx7tcompare})(b),(c)). 
Thus $\lambda = 0$ does not scramble. For the $\lambda = 0.3$ and $\lambda = 0.8$ we observe 
operator spreading at the {\it maximal} group velocity for $|x|<\xi_{OTOC}$ (where
$\xi_{OTOC}$ characterises a length sufficiently large compared to the
localization length). However, for values of $|x|>\xi_{OTOC}$ 
$C(x,t)=0$ for all times. $\xi_{OTOC}$ is shown in Fig.~\ref{fig:OTOClam}(b),(c) as 
the dashed vertical red lines and indicated the length scale beyond which $C(x,t)<10^{-3}$ for all times.
Hence, the operator radius is bounded by $\xi_{OTOC}$ and does not spread into regions beyond $\xi_{OTOC}$.
As expected, $\xi_{OTOC}$ shrinks with increasing $\lambda$, as seen in Fig.~\ref{fig:OTOClam}(b),(c).
For $|x|<\xi_{OTOC}$, $C(x,t)$ initially grows with $t$ until it peaks and then decreases to weakly oscillate around
a non-zero value, and never returns to zero. This is a fundamentally different behaviour
than the no disorder case. This long-time limit of $C(x,t)$ for $|x|<\xi_{OTOC}$ increases weakly with $\lambda$
while it decreases with $x$. The light cone has therefore the shape of a neck-tie with a v-shaped tip. This behaviour is markedly different from results in MBL systems where a much different logarithmic lightcone has been observed~\cite{Luitz2017,Deng2017,Sahu2018}.
	
\begin{figure}[!ht] \centering
		
                \includegraphics[width=\linewidth]{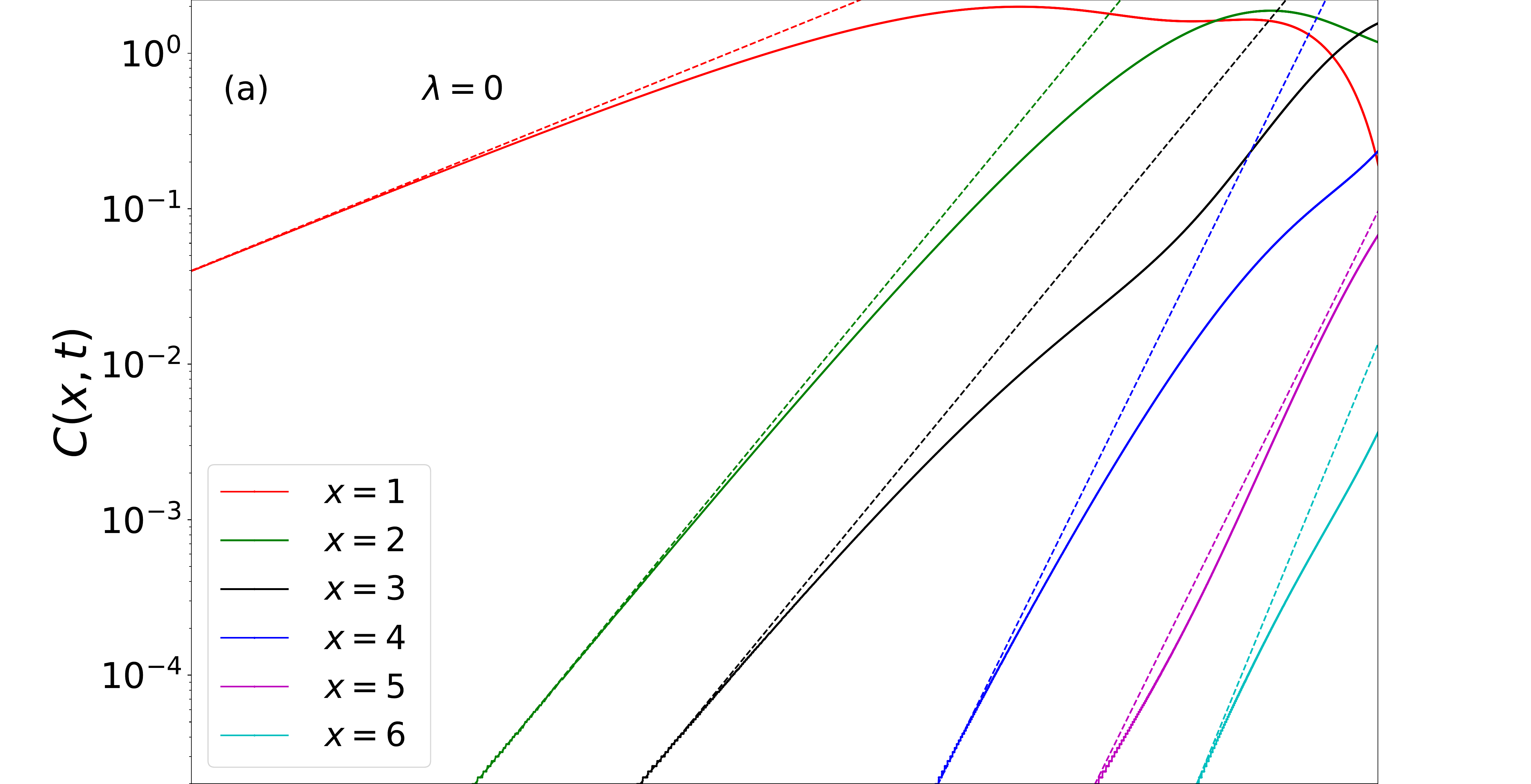}
	
                \includegraphics[width=\linewidth]{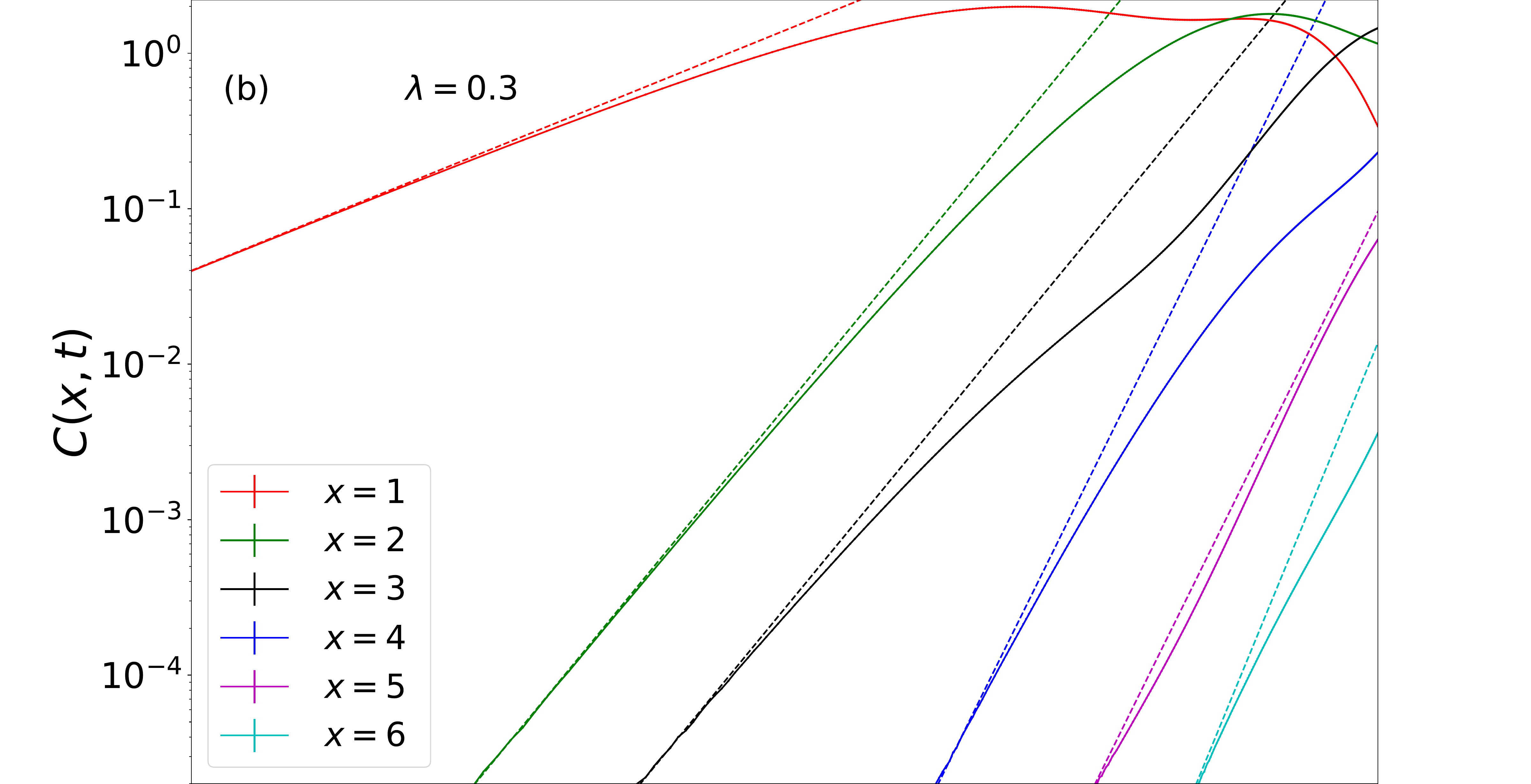}
			
                \includegraphics[width=\linewidth]{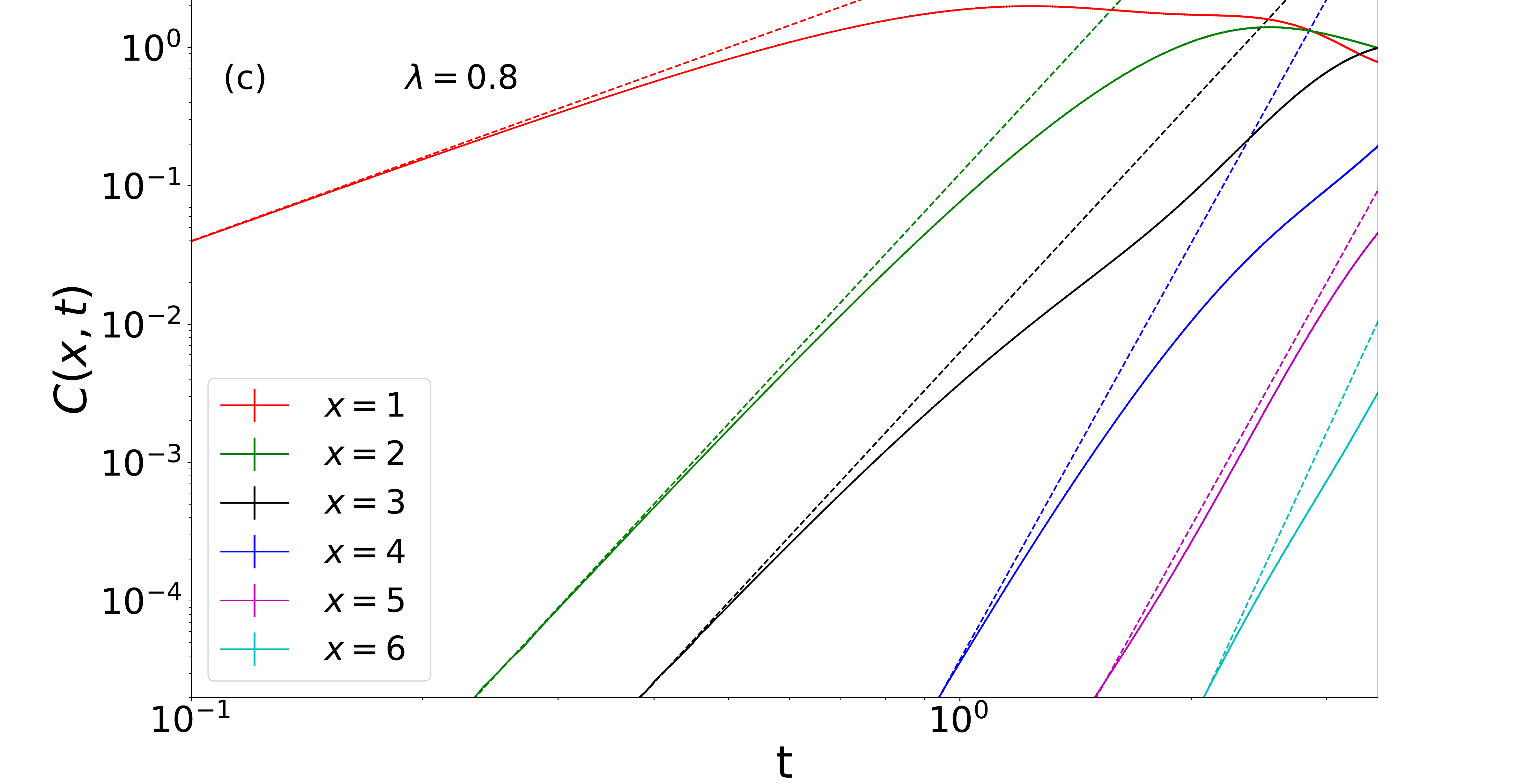}
	
\caption{Early time $C(x,t) $ at different values of $x$ for
each studied $\lambda$. The dotted lines for $x=1,3,5,6$ are
the power laws $t^{2|x|}$ with appropriate constants in front
while the solid lines are the actual data. For
  $x=2,4$ the next leading order power, $t^{2(|x|+1)}$, is
required to fit the data for every value of $\lambda$.}
\label{fig:prodearly} 
\end{figure}

In Ref. \cite{Hamza2012,Huang2017} it has been noted that the Anderson localized states do
exhibit a non-expanding light-cone with the commutator between two operators
being bounded in time by, 
\begin{equation}
		|| [A(0,0), B(x,t)]|| \leq C e^{-\frac{|x|}{\xi}},
\end{equation} 
where $A(0,0)$ and $B(x,t)$ are operators with local support and $x$ is the
displacement in between them. This result implies that $C(x,t)$ should have the same
exponential behaviour and we have verified that the results in Fig.~\ref{fig:cx7tcompare}(a) for $\lambda=0.8$
and for $|x|<8$ is well described by:
\begin{equation}
  C(x,t=64)\sim e^{-a|x|},
\end{equation}
with $a\sim 0.33$.
In Ref. \cite{LinOTOCising} it was proposed that a universal power law, applies
to all lattice systems where the Hamiltonian is constructed from local
interactions. For the quantum Ising model this was shown to be~\cite{LinOTOCising} $C(x,t)\sim t^{2(2x-1)}$.
This is seen by considering the Hadamard formula, and an operator
$\hat{A}$ (see ref. \cite{wmillersymmetry} lemma 5.3). 
\begin{equation}
e^{s \hat{H}} \hat{A}e^{-s\hat{H}} = \hat{A} + s[ \hat{H}, \hat{A} ] + \frac{s^2}{2!} [ \hat{H},[\hat{H},\hat{A}]] \dots = \sum_{n=0}^\infty \frac{s^n}{n!}L_n.
\end{equation}
For the RFXX spin chain considered here we arrive at a slightly modified power-law by repeating the argument of Ref.~\cite{LinOTOCising}. This is done by considering
$\hat{A} = \hat{\sigma}_{\frac{L}{2}}^z$ and $s = it$ and determining the smallest $n$ of the above sum, such that $[L_n,\sigma_{j=x}^z] \neq 0$. 
This corresponds to successively evaluating the commutator between the
Hamiltonian and the string of operators that grows until it reaches $j=x$. The
strings which appear at the smallest order of $t$ for odd $n$ look like
(shifting the indexes for simplicity),
$\hat{\sigma}_0^x\hat{\sigma}_1^z\hat{\sigma}_2^z \dots \hat{\sigma}_{x-1}^z
\hat{\sigma}_x^y$ and $\hat{\sigma}_0^y\hat{\sigma}_1^z\hat{\sigma}_2^z \dots
\hat{\sigma}_{x-1}^z \hat{\sigma}_{j=x}^x$ while for $n$ even,
$\hat{\sigma}_0^x\hat{\sigma}_1^z\hat{\sigma}_2^z \dots \hat{\sigma}_{x-1}^z
\hat{\sigma}_x^x$ and $\hat{\sigma}_0^y\hat{\sigma}_1^z\hat{\sigma}_2^z \dots
\hat{\sigma}_{x-1}^z \hat{\sigma}_{j=x}^y$, yielding $n=j=x$. At least for regions inside the
light-cone we expect this behaviour to be independent of $\lambda$. With $C(x,t)$ the square
of the commutator we then find for the RFXX model at early times,
\begin{equation} \label{eq:powerlaw}
		C(x,t) \sim  t^{2|x|},
\end{equation}
with a power law that is independent of $\lambda$ and is therefore {\it not} modified by the presence of disorder.
This is purely a quantum mechanical phenomenon occurring before the wave front hits and is not a signature of scrambling. 
This phenomenon is captured in Fig. \ref{fig:prodearly} where results are shown for $\lambda=0$ (Fig.~\ref{fig:prodearly}(a)), $\lambda=0.3$ (Fig.~\ref{fig:prodearly}(b))
and $\lambda=0.8$ (Fig.~\ref{fig:prodearly}(c)) where results are shown for a range of values of $x$ confirming the above power-law dependence.
For $|x|=2,4$ we include the next leading term in the fits: $t^{2(|x|+1)}$.
The power law growth in this model is thus universal for $\lambda=0$
as well as in the the localized phase ($\lambda\neq 0)$, assuming we are inside the light-cone. Interestingly
outside of the light-cone, despite the derivation for the power law being
independent of $\lambda$, the power law breaks down, signifying localization
suppressing quantum effects as well. Precisely, how localization effects will start to dominate is not clear, although the clear presence of
correction terms for $|x|<\xi_{OTOC}$ are an indication that such corrections eventually become dominant.

\begin{figure}[!ht]
\centering
\includegraphics[width=\linewidth]{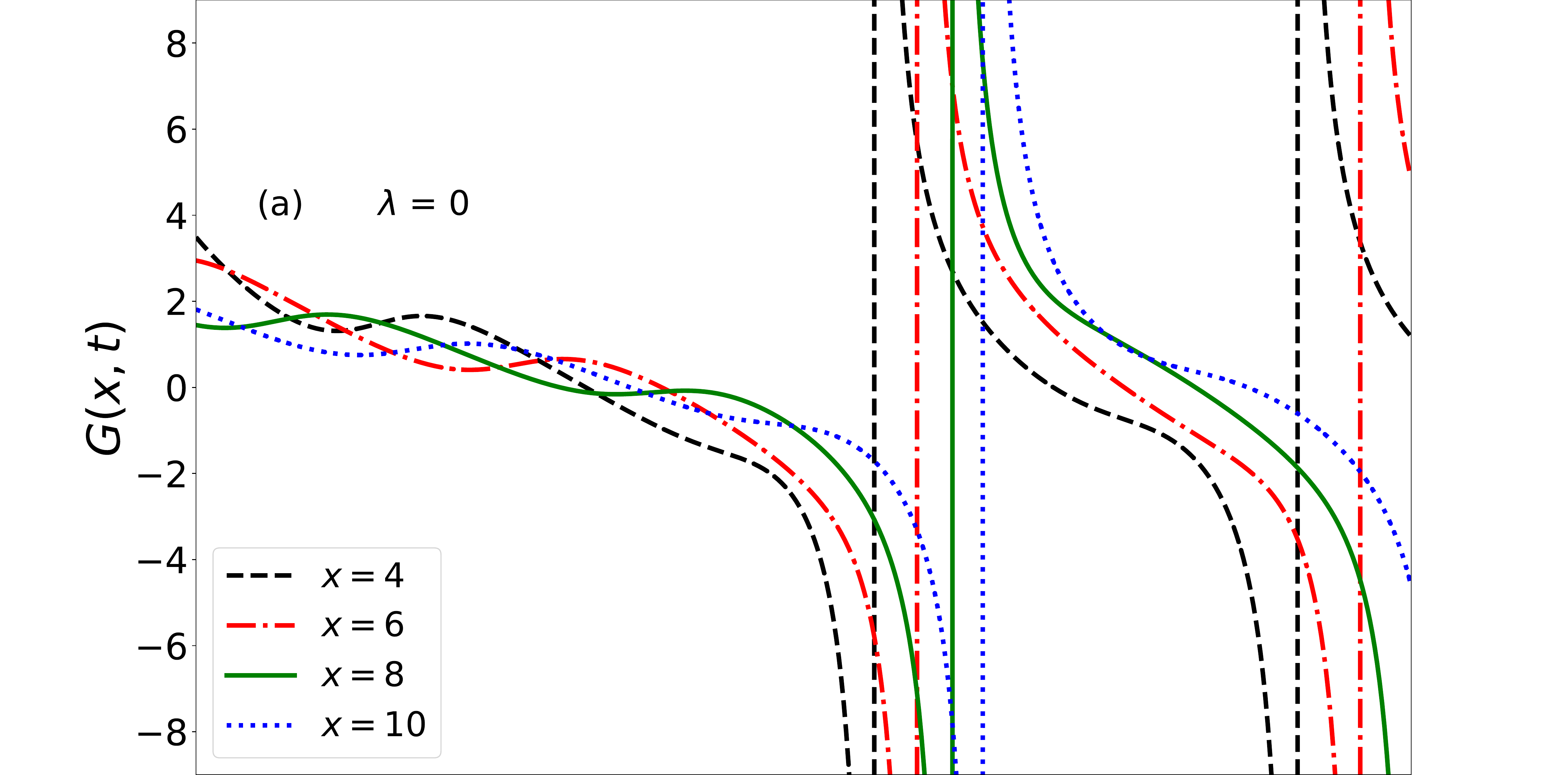}

\includegraphics[width=\linewidth]{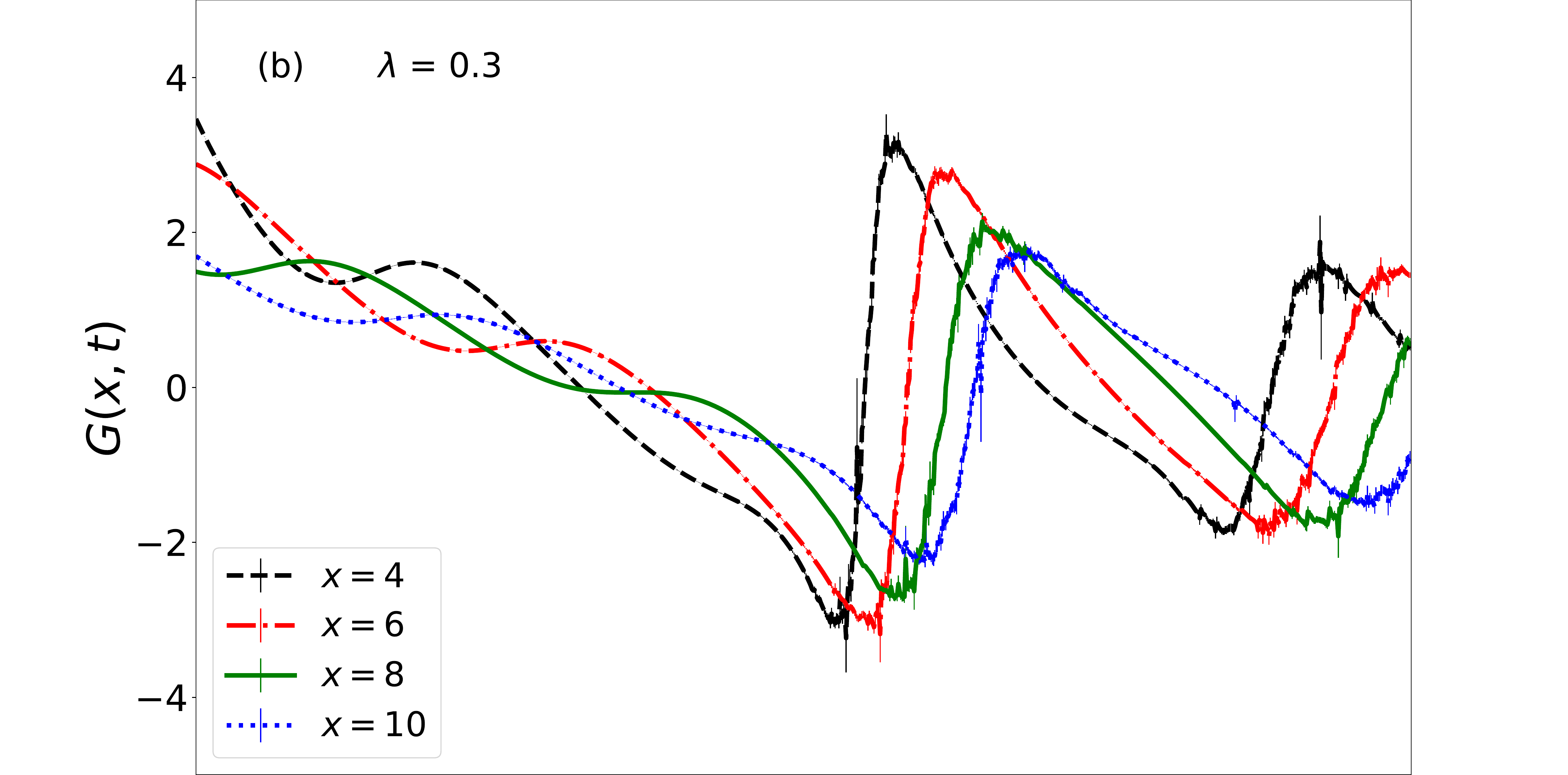}
		
\includegraphics[width=\linewidth]{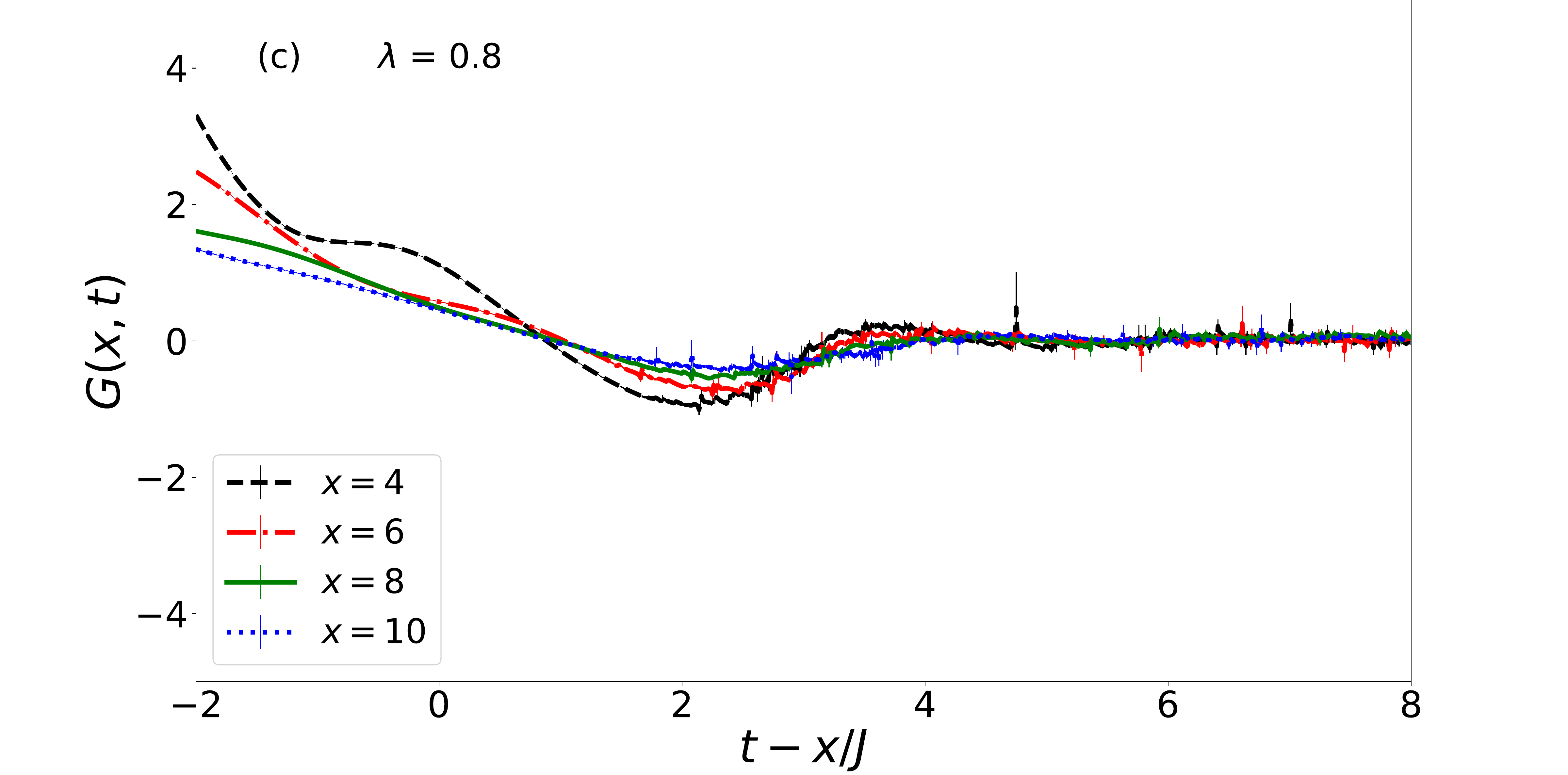}
\caption{$G(x,t)$ graphed against $t-\frac{x}{t}$. This simulation required 5,000 realizations of the random Hamiltonian to get reasonable error bars.}
\label{fig:wavefront}
\end{figure}

Finally we study the behaviour of $C(x,t)$ at the wave front which moves at a
velocity $v_{max} = J = 1$. Here, we use (following Ref. \cite{LinOTOCising}) 
the function, 
\begin{equation}
G(x,t) = \frac{\partial \ln C(x,t)}{\partial t} = \frac{1}{C(x,t)} \frac{\partial C(x,t)}{\partial t}.
\end{equation}
Since we know the expression for $C(x,t)$ exactly $G(x,t)$ can be calculated without resorting to 
evaluating the derivatives numerically. Our results for this function are plotted in Fig. \ref{fig:wavefront}. 
The wave front hits when $t-\frac{x}{J} = 0$, and we again see the initial purely quantum mechanical growth of $C(x,t)$ before the front hits. 
After the wave front hits $G(x,t)$ in all cases becomes negative after a short time, and then an oscillatory behaviour about $0$ is observed. 
For $\lambda = 0$, the repeating pattern appears to have a discontinuous change when
going from negative to positive values of $G(x,t)$, however this is most likely an artifact of 
$C(x,t)$ returning to zero and bouncing back upwards as seen in Fig. \ref{fig:OTOClam}. 
Because this behaviour is observed for extremely large values of 
$t$ and large accessible system sizes, we cannot conclude exactly how $C(x,t)$ goes to zero as $t\to\infty$ for $\lambda=0$.  
For $\lambda\neq 0$ the behaviour is different since $C(x,t)$ does not go back to zero, but instead oscillates around a non-zero value. 
However, we see that as $\lambda$ is increased $G(x,t)$ varies much less rapidly. 
Both $\lambda=0.3$ and $0.8$ show oscillatory behaviour in $G(x,t)$ after the wave front reaches but the amplitudes are suppressed with larger $\lambda$. 
Interestingly, we do not observe monotonic behaviour on any meaningful interval.

\subsection{Thermal States}
\begin{figure}[htb]
\centering
	
\includegraphics[width=\linewidth]{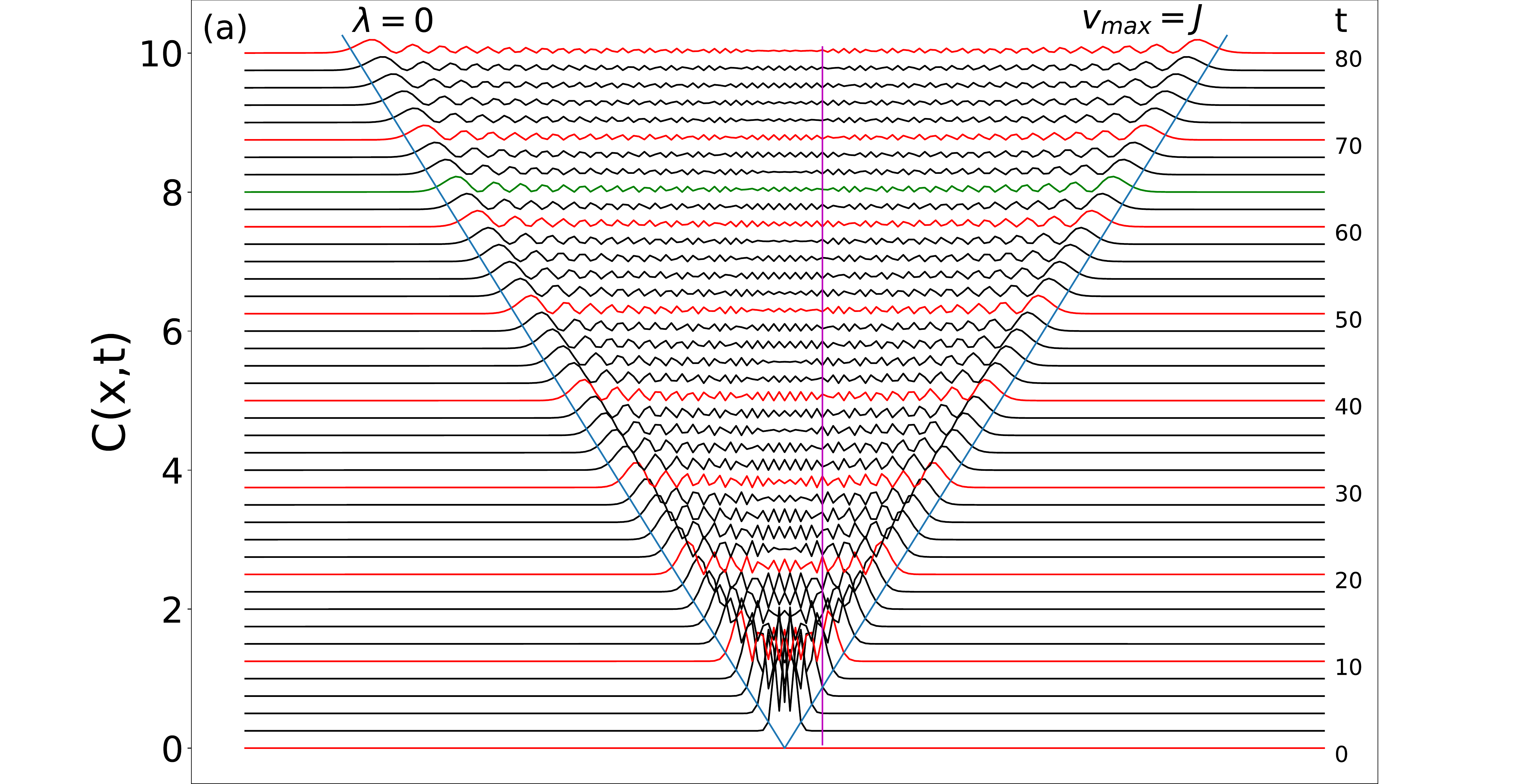}
		
\includegraphics[width=\linewidth]{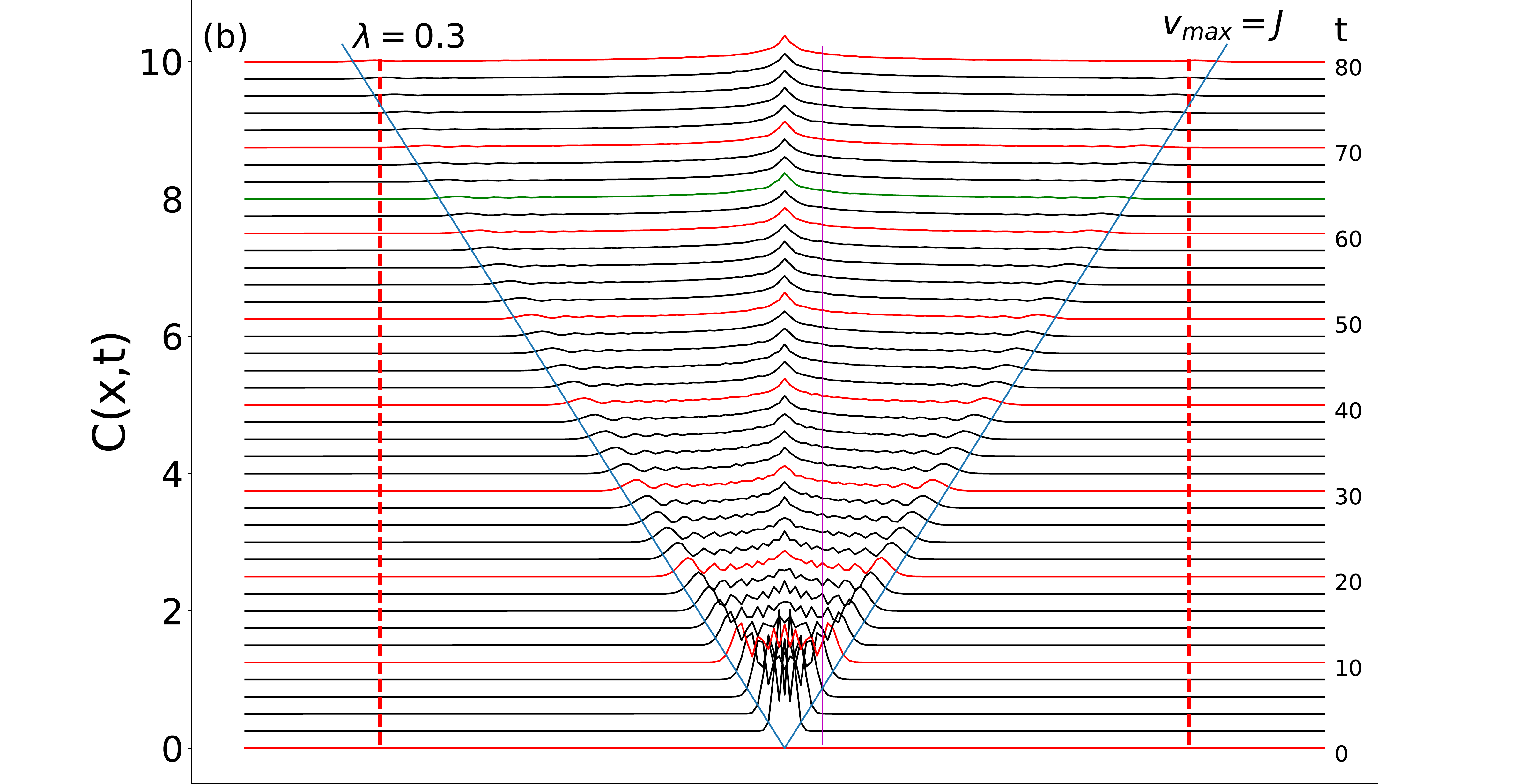}
		
\includegraphics[width=\linewidth]{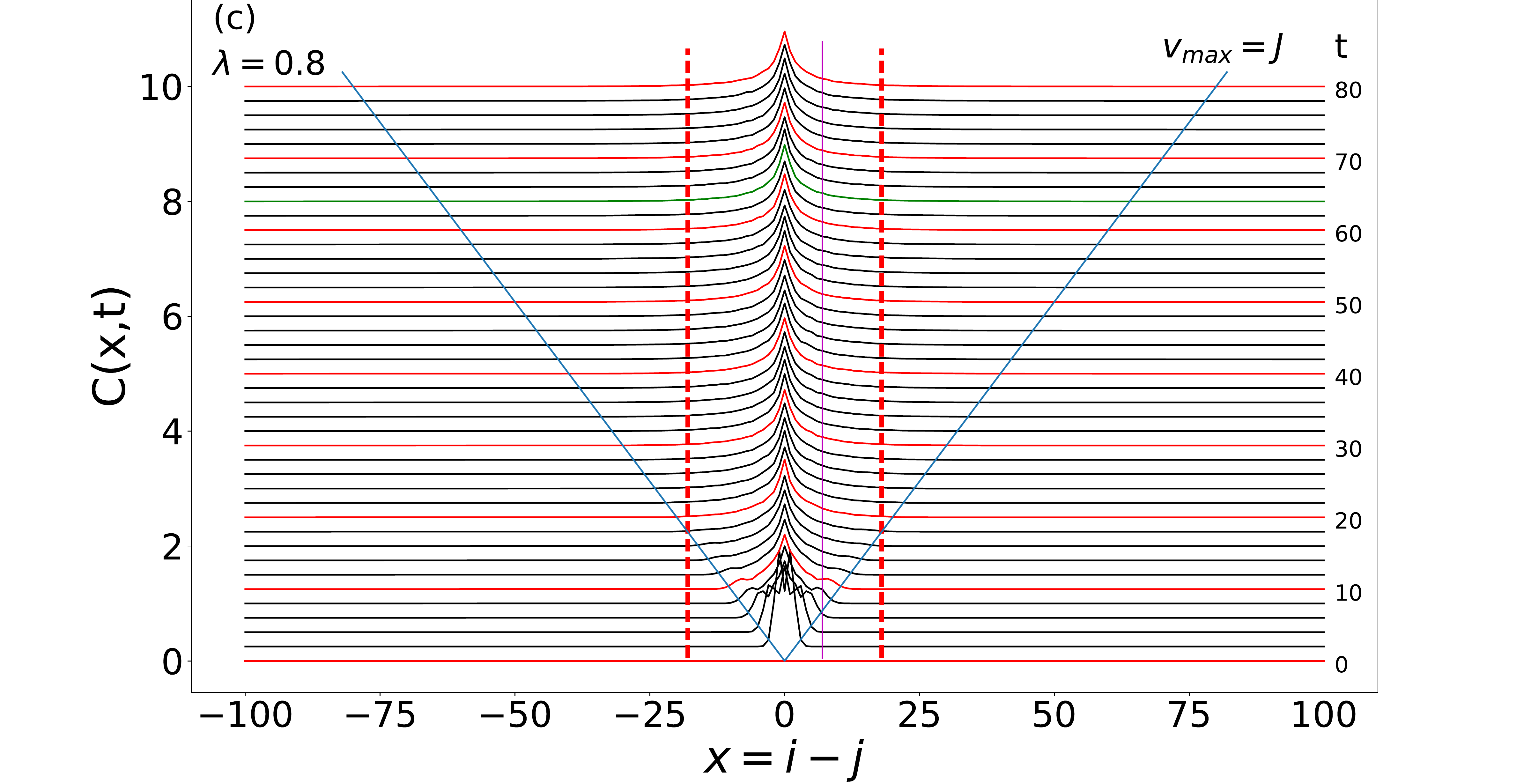}

\caption{Wave propagation plot of $C(x,t)$ for the RFXX spin chain at disorder
  strength (a) $\lambda = 0$, (b) $\lambda=0.3$ and (c) $\lambda=0.8$ in a thermal state with $\beta=1$.
  The $x$-axis is the displacement from the center of the chain $i=\frac{L}{2}$.
  symmetry about the position $i$ the wave propagates symmetrically. The two
  $y$-axis are the values $C(x,t)$ and the corresponding time. The maximal
  group velocity $v_{max} = J$ is also shown as the solid blue line.
  In panel (b) and (c)
  the vertical dashed red line indicates $\xi_{OTOC}$,  the $x$ value beyond which $C(x,t)< 10^{-3}$
  for any $x$. $\xi_{OTOC}= 18$ for $\lambda = 0.8$ and 75 for $\lambda =0.3$.
  } \label{fig:thermOTOClam}
\end{figure}

Next we repeat these calculations, but with a thermal state with $\beta=1$ instead of the product state considered in the previous section. 
$\beta = 1$ is an arbitrary choice because the dynamics will overall depend primarily on the anti-commutator in time (which is $\beta$ independent), for both disorder and non-disorder.
Hence, the variation with $\beta$ is relatively minor in particular in the presence of disorder. 
This state is already in equilibrium and exhibits a significantly different expression for $C(x,t)$ as detailed in Eq.~(\ref{thermF}). 
In Fig. \ref{fig:thermOTOClam} we show $C(x,t)$ at different time slices. Although this plot looks similar to the product state version, Fig.~\ref{fig:OTOClam}, differences emerge.
Firstly the peaks of the $C(x,t)$ are smaller than was the case for the  product state, and the $C(x,t)$ is much smoother as seen in Fig.~\ref{fig:cx7tcompare}, 
travelling simply as a smooth parabola like curve in space. However, the oscillatory behaviour occurs also in this case, and we again do not expect to be able to 
find a description for how $C(x,t)$ approaches zero in late time. 
For this value of $\beta=1$ we find the same values for $\xi_{OTOC}$ as was determined for the N\'eel product state.

\begin{figure}[htb]
\centering
\includegraphics[width=\linewidth]{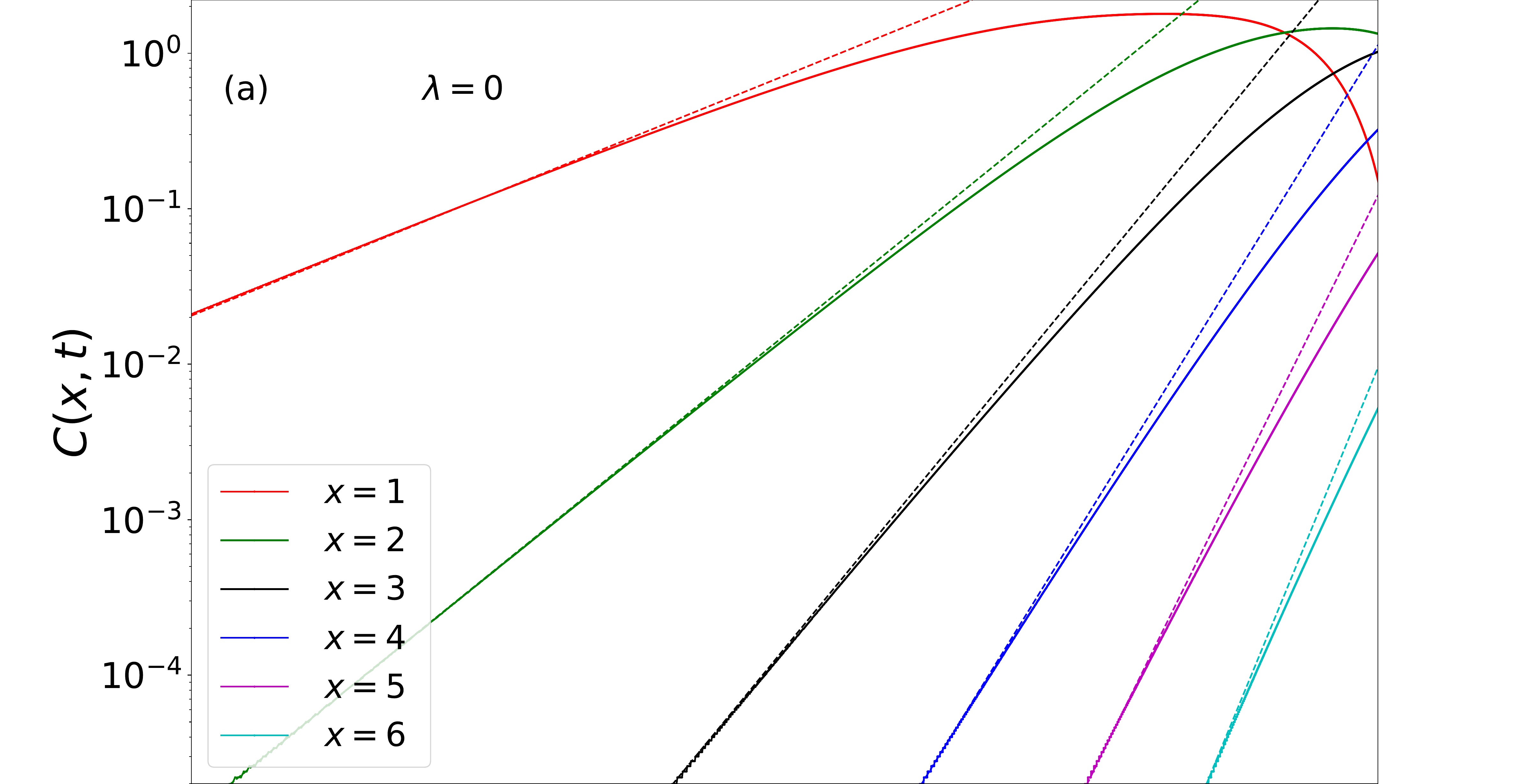}
	
\includegraphics[width=\linewidth]{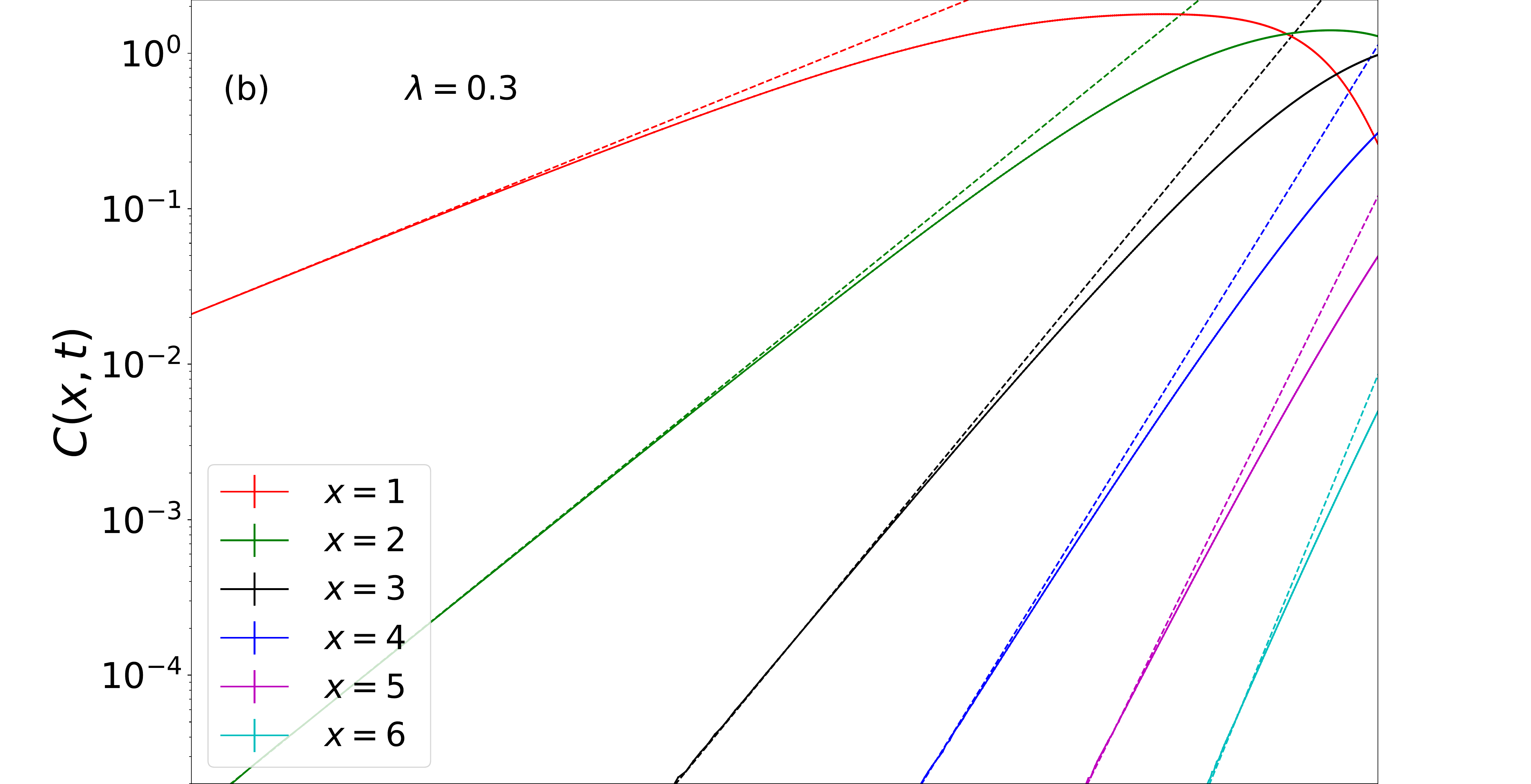}

\includegraphics[width=\linewidth]{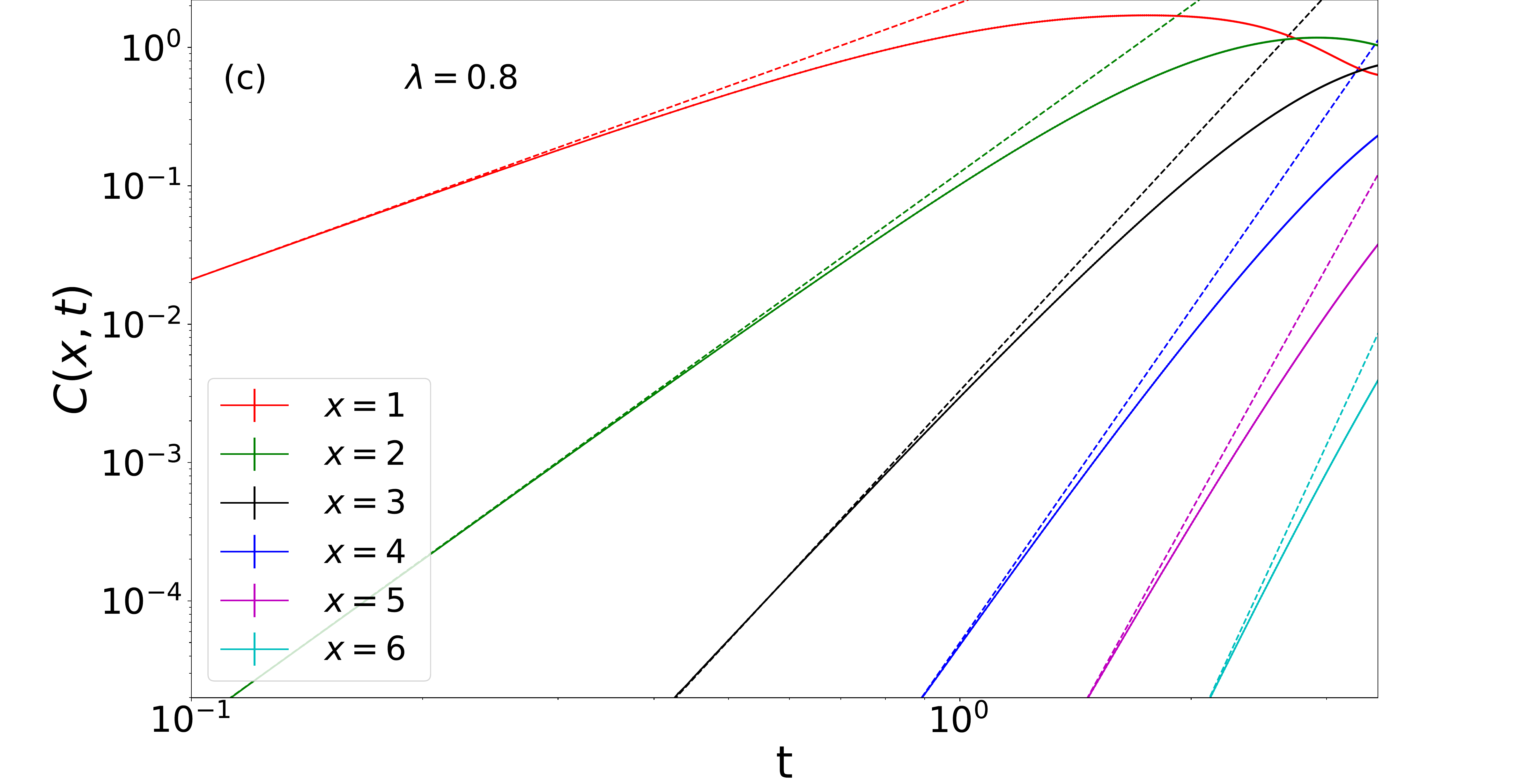}

\caption{Early time $C(x,t) $ thermal correlations at different values of $x$
  for $\lambda=0,0.3,0.8$. The dotted lines for $x=1,2,3,4,5,6$ are the
  power laws $t^{2|x|}$ with appropriate constants in front while the solid lines
  are the actual data. }
\label{fig:thermearly}
\end{figure}
We also see in Fig. \ref{fig:thermearly} that the thermal states obey the power
law discussed in Eq. (\ref{eq:powerlaw}). For the thermal state the agreement with the power-law
behavior is better than for the product state and
no higher order terms are included in the fits shown in Fig.~\ref{fig:thermearly}. This
is most likely due to the absence of noise, which indicates modelling the
wavefront will be easier with this initial condition.
\begin{figure}[!ht]
\centering
		
\includegraphics[width= \linewidth]{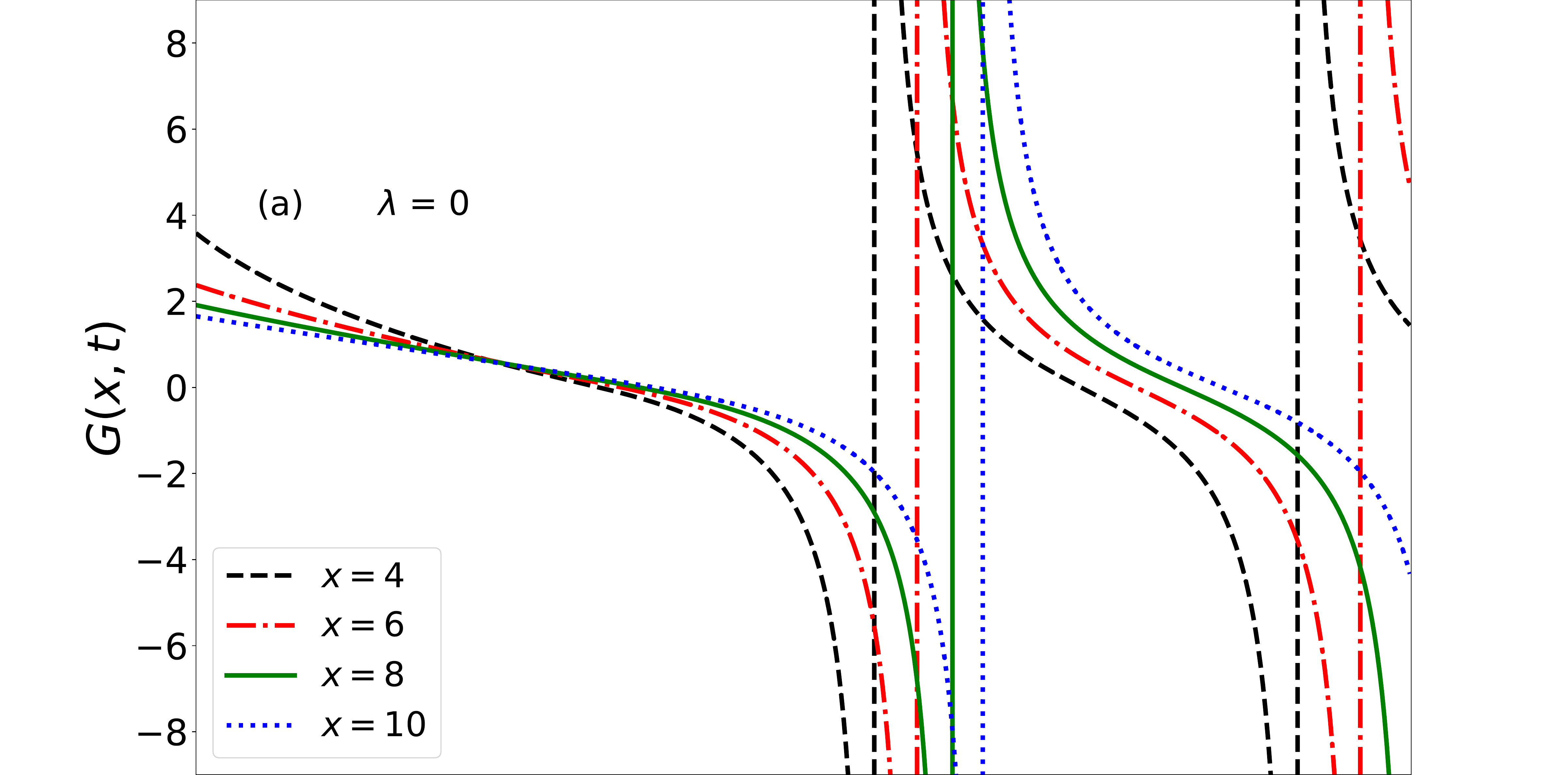}
		
\includegraphics[width= \linewidth]{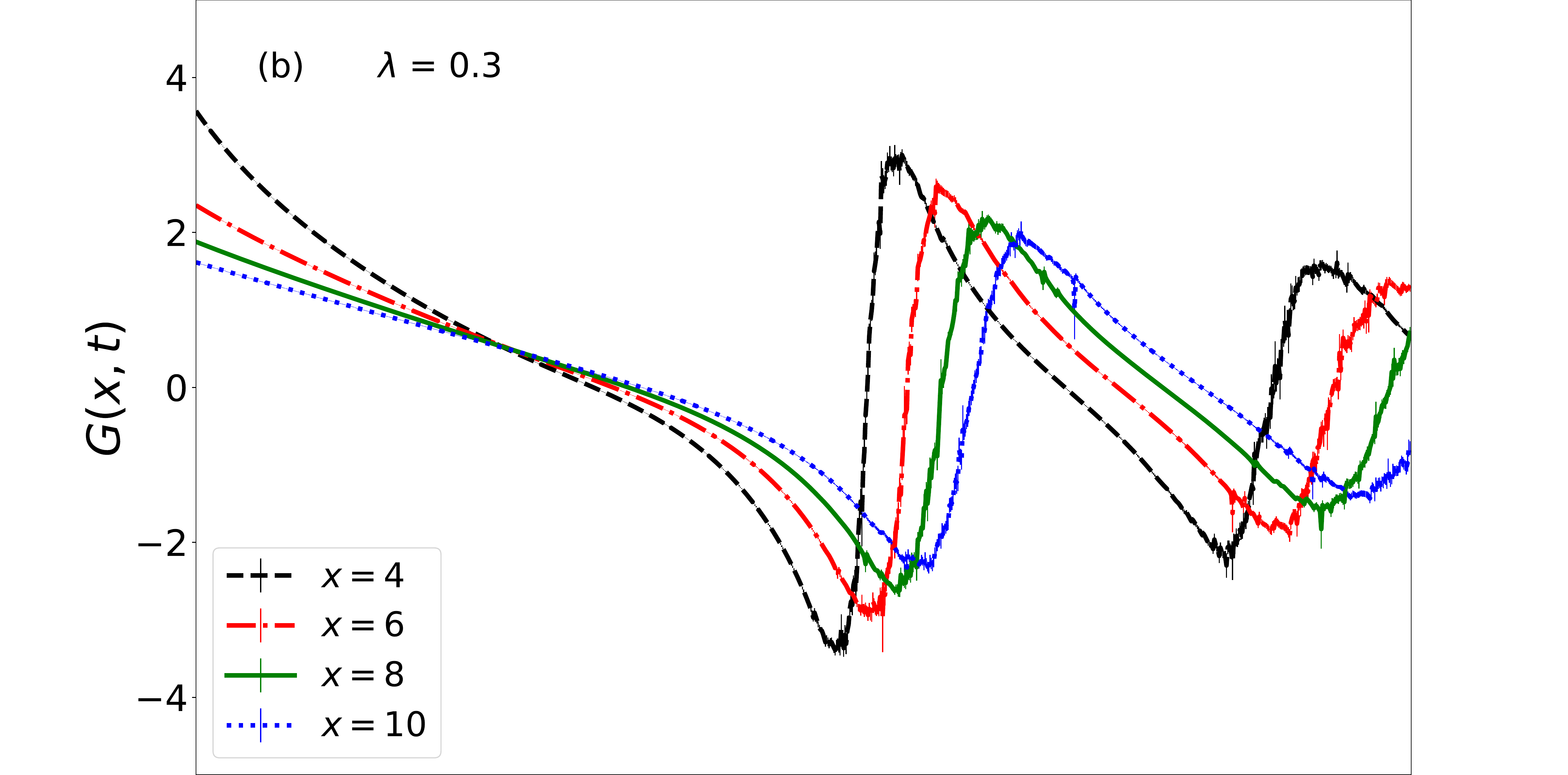}
		
\includegraphics[width= \linewidth]{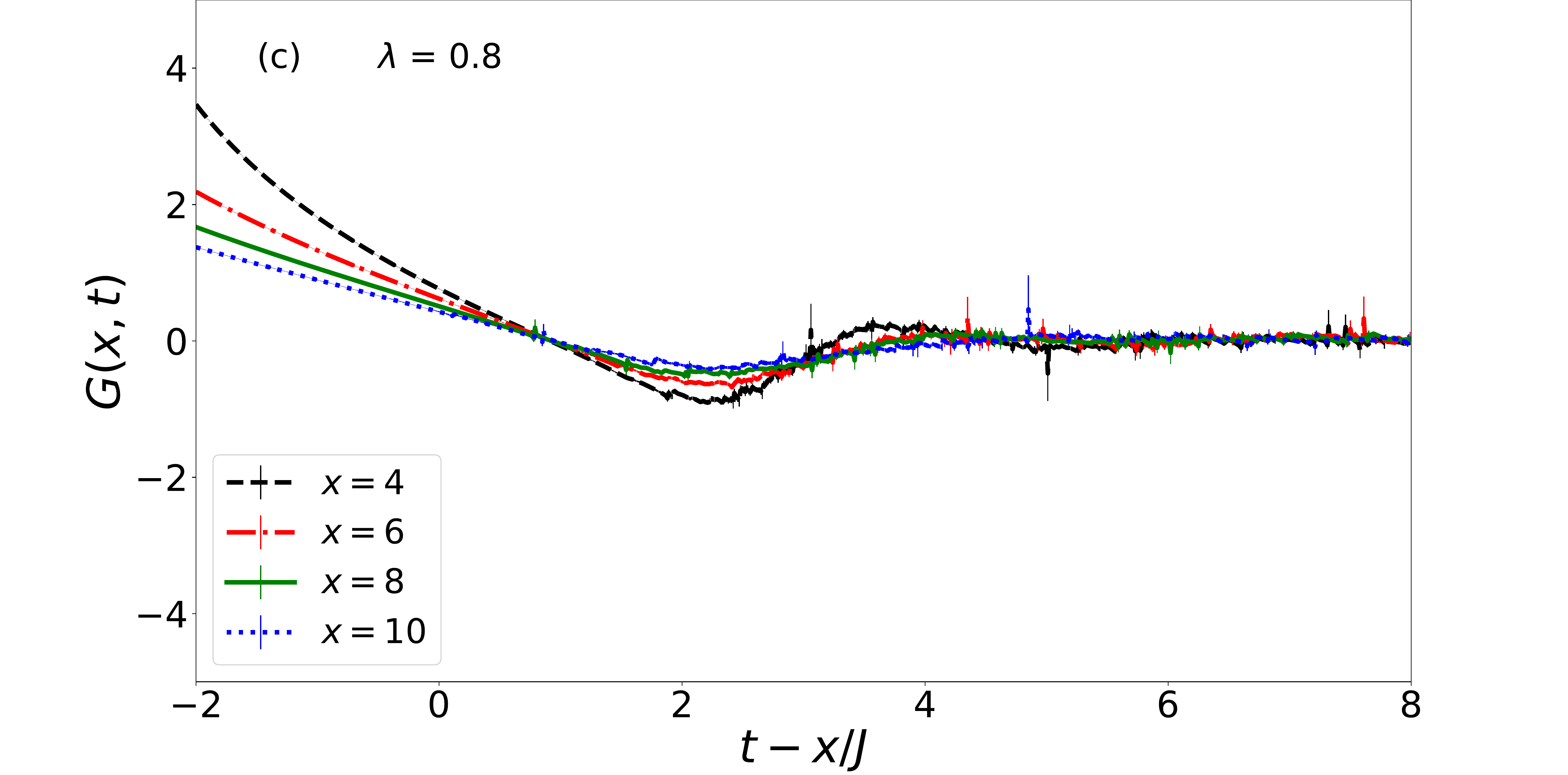}
	
\caption{$G(x,t)$ as a function of $t-\frac{x}{J}$ calculated in the thermal state with $\beta=1.$
This simulation required 5,000 realizations of the random
Hamiltonian to get reasonable error bars.} \label{fig:thermwavefront}
\end{figure}
	
Finally, in Fig. \ref{fig:thermwavefront}, we show the wavefront as described
by $G(x,t)$ evaluated using the thermal state with $\beta=1$.  Unlike the
product state we observe monotonic behaviour for the approximate region
$t-\frac{x}{J} \in [-2,2]$ and we observe strong $x$ and $\lambda$ dependence.
Once again the $\lambda = 0$ diverges when $C(x,t)$ goes to zero, and the
$\lambda\neq 0$ cases do not exhibit this behaviour due to $C(x,t)$ never returning
to zero. Similarly, we observe oscillatory behaviour after the wavefront passes.  At
the wave front which we define as $t- \frac{x}{J}\in[0,2]$ we can effectively
approximate $G(x,t)$ by a linear equation $G(x,t) \approx  m(t-x/J)t+c = at+b$, due to the
shapes of the functions we expect $a = a(x,\lambda)$ and $b = b(x,\lambda)$. 
Interestingly this form suggests that at the wavefront, 
\begin{equation} \label{eq:wavefront}
		C(x,t) \sim e^{\frac{a(x,\lambda)t^2}{2}+b(x,\lambda)t}.
\end{equation}
To follow the universal form of Eq.~\ref{eq:uniC} one must have 
\begin{equation} \label{eq:uniG}
		G(x,t) \sim \frac{\lambda_L}{t^{p+1}}(x-v_Bt)^p(v_Bt+px).
\end{equation}
However, the form of Eq.~(\ref{eq:uniG}) does not permit a linear equation. Thus we conclude that our results in Eq.~(\ref{eq:wavefront})  do not follow the proposed universal form, Eq.~(\ref{eq:uniC}).
We currently do not know an exact expression for $a(x,\lambda)$ and $b(x,\lambda)$, however, for completeness we provide a
table of the fitted values in table \ref{table}.  The values for $a=m$ are necessarily
negative and $c$ positive. The errors reported are one standard derivation. The
small errors indicate that the form given in Eq.~(\ref{eq:wavefront}) is a
reasonable description. 
\begin{table}[htb] 
\centering
\begin{tabular}{r|c|c|}
\centering
$\lambda = 0$ & $m$ & $c$   \\
\hline
$x=2$ & -0.72948875 $\pm$ 0.002 & 0.94258389 $\pm$ 0.002  \\
\hline
$x=4$ & -0.59368064 $\pm$ 0.001 & 0.88059725 $\pm$ 0.001 \\
\hline
$x=6$ & -0.5040499 $\pm$ 0.0009 & 0.82791869 $\pm$ 0.001  \\
\hline
$x=10$ & -0.44021805 $\pm$ 0.0008 & 0.78445194 $\pm$ 0.0009 \\
\hline
$\lambda = 0.3$ & $m$ & $c$   \\
\hline
$x=2$ & -0.73975833  $\pm$ 0.002 & 0.92284197  $\pm$ 0.002  \\
\hline
$x=4$ & -0.60431888  $\pm$ 0.001 & 0.85326093  $\pm$ 0.001  \\
\hline
$x=6$ & -0.51342755  $\pm$ 0.001 & 0.7949133  $\pm$ 0.001  \\
\hline
$x=10$ & -0.44742463  $\pm$ 0.0009 & 0.74793077  $\pm$ 0.001 \\
\hline
$\lambda = 0.8$ & $m$ & $c$   \\
\hline
$x=2$ & -0.81713901  $\pm$ 0.002 & 0.74741784  $\pm$ 0.003 \\
\hline
$x=4$ &-0.63429921  $\pm$ 0.002 & 0.60081936  $\pm$ 0.003 \\
\hline
$x=6$ & -0.52336579  $\pm$ 0.003 &  0.4942889  $\pm$ 0.003  \\
\hline
$x=10$ & -0.40722873  $\pm$ 0.003 & 0.40339502  $\pm$ 0.004 \\
\hline
\end{tabular}
\caption{
Results of fitting the function  $G(x,t) \approx m(t-x/J)t+c = at+b$ where  on the interval  $t-
\frac{x}{J}\in[0,2]$ for different values of $\lambda$ and $x$. The errors
reported are one standard deviation on the parameter.   }
\label{table} 
\end{table}
\\

\section{Bipartite entanglement entropy}
\label{Ent}
We now turn to a discussion of the growth of entanglement in the RFXX starting from the N\'eel product state which, due to its product form, has zero entanglement.
The entanglement entropy between two subsystems $A,B$ is defined with the reduced density matrices $\rho_A = \tr_B \rho$ and $\rho_B = \tr_A \rho$, 
\begin{equation}
		S_{A,B} = -\tr \left( \rho_A \ln \rho_A \right) = -\tr \left( \rho_B \ln \rho_B \right).
\end{equation}
Where the equality is taken because regardless of the partition $\rho_A$ and
$\rho_B$ have identical non-zero eigenvalues \cite{PeschelEE}.  For the
remainder of this section we partition the lattice into halves and denote this
quantity as $S_{\frac{L}{2}}$. 
	
Rigorous bounds for the entanglement entropy in the RFXX model in the Anderson localized phase have been derived and it is expected to obey an area law in one dimension \cite{Pouranvari2015,Abdul-Rahman2016,RahmanXY}. 
In particular, it has been shown that the growth of entanglement remains bounded for all times~\cite{Abdul-Rahman2016}.
This means entanglement entropy even for arbitrarily small disorder strengths will be bounded by a constant in the late time limit. The approach to this limiting
value is relatively less explored and that is our focus here. Exact diagonalization results on small systems have been discussed in Ref.~\cite{BardarsonEnt} where for relatively strong disorder
the entanglement entropy reached a constant at very short times.

In order to study the time dependent entanglement we time evolve our state,
Eq.~(\ref{prodstate}) and calculate the entanglement entropy at late times.  We
expect that at sufficiently large system sizes we will not observe an increase
in entanglement entropy as the system grows, since we will be close to the
theoretical maximum.  In Ref.~\cite{BardarsonEnt} the authors did a similar
calculation for both Anderson and many body localized phases.  However
comparing many body localized systems to Anderson localized systems restricts
the system sizes, here we do not have this restriction, focusing entirely on
the Anderson localization regime.  Using the method in \cite{LatorreEE} we can
efficiently calculate the entanglement entropy from the occupation matrix
defined in Eq.~(\ref{occmatrix}). Since we are interested in late time
entanglement entropy, it is tempting to consider the infinite time average of
the occupation matrix. That is, for each element, we define (similar to
\cite{timeevolvefermions}),
\begin{eqnarray}
\Lambda^f(\infty)_{i,j} := \lim_{T \to \infty} \int_{0}^T dt \frac{1}{T} \langle \hat{f}_i^\dagger (t) \hat{f}_j(t) \rangle  =  \nonumber\\
\lim_{T \to \infty} \int_{0}^T dt \frac{1}{T} \sum_{k,l}  e^{i(\epsilon_k-\epsilon_{l}) t}A_{i,k} A_{j,l} \langle \hat{d}_k^\dagger \hat{d}_l \rangle \nonumber\\
= \sum_k A_{i,k}A_{j,k} \langle \hat{d}_k^\dagger \hat{d}_k \rangle.
\end{eqnarray}
Which amounts to a "dephasing" of the off-diagonal contributions. 
Note that we used the fact that the $\epsilon_{k}$ are expected to be non-degenerate \cite{StolzALintro}.  
The infinite time average occupation matrix corresponds to a generalized Gibbs ensemble, 
\begin{equation}
\rho = \frac{1}{Z}e^{-\sum_k \beta_k \hat{Q}_k},
\end{equation}
where $\hat{Q}_k = \hat{d}_k^\dagger \hat{d}_k$. 
	
\begin{figure}[!ht]
\centering
\includegraphics[width=\linewidth]{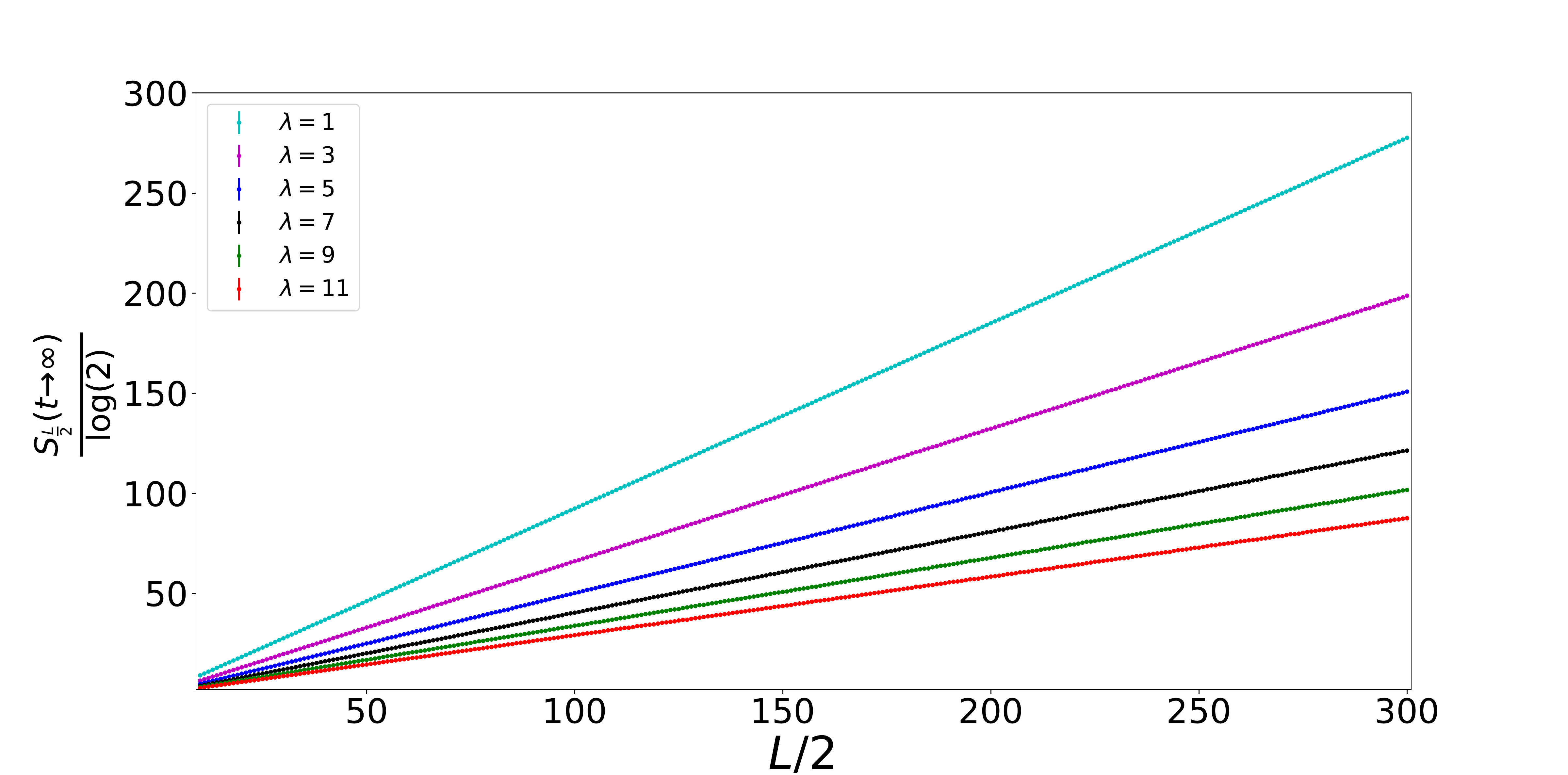}
\caption{
Infinite time average $S_{\frac{L}{2}}$ plotted against system size. Each point
is an average over $5000$  random field realizations and the error shown is
the standard error on the calculated mean. System sizes are taken from $L=20$
to $L=600$. For these results the approximation yielding the infinite time average
  is {\it not} valid and the resulting volume law is {\it incorrect}}
\label{fig:inftimeent}
\end{figure}
However in Fig.~\ref{fig:inftimeent} we see that the infinite time average
occupation matrix predicts volume laws despite large disorder, the disorder
only changes the slope, but the entanglement entropy still grows linearly with
system size. This is most likely due to the infinite time average being a valid
approximation for the equilibrated occupation matrix {\it only on small sub-systems}, where
the difference disappears with the system size. However, here we are focusing on sub
systems which are a constant fraction of the system we are growing. Thus the
errors that disappear on a small scale add up on the macroscopic scale and we
lose the ability to effectively describe the equilibrated state with the
infinite time average. The above approximation is therefore {\it not valid} in the present case.
Instead we must pick an arbitrary late time to calculate
the entanglement entropy which we here take to be $t=10^{11}$.

\begin{figure}[!ht]
\centering
\includegraphics[width=\linewidth]{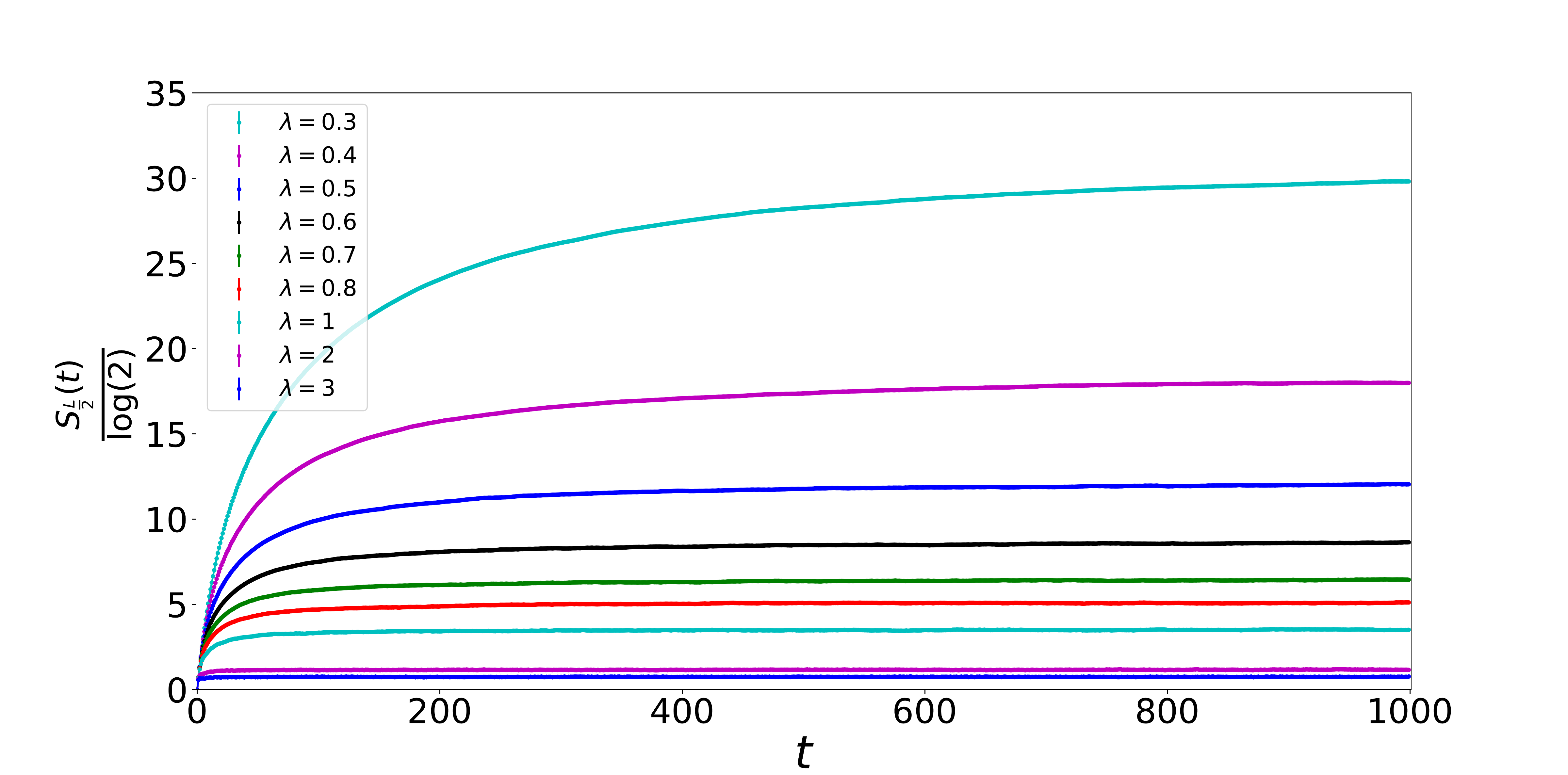}

\caption{ $S_{\frac{L}{2}}$ plotted against time for a system size of $L =
  400$. Results are shown for $\lambda=0.3,0.4,0.5,0.6,0.7,0.8,1,2,3$. Each point is an average over $1000$ field realizations and the error
  shown is the standard error on the calculated mean. }
\label{fig:entropyvstime}
\end{figure}

In Fig. \ref{fig:entropyvstime} we show results for  the growth of the average entanglement entropy 
with time for the range of disorders we are interested in. It has been proposed that the saturation time
for entanglement entropy  $\log (t_{sat}) \sim L$~\cite{BardarsonEnt}. 
Intuitively, for the Anderson insulator, taking a localization value
$\xi(\lambda) \ll L$ we would expect the time it takes for the entanglement
entropy to get close to this saturated value to be much smaller, as only small
subsystems become entangled with each other. This is indeed what we see in Fig.
\ref{fig:entropyvstime}, by $t=500$ all but $\lambda = 0.3$ have little to no
growth, and $\lambda = 0.3$ has slowed significantly compared to its initial
rise. However, the approach to a constant value could involve logarithmic factors and for subsequent analysis we 
therefore chose to study the entanglement at $t=10^{11}$.
\begin{figure}[!ht]
\centering
\includegraphics[width=\linewidth]{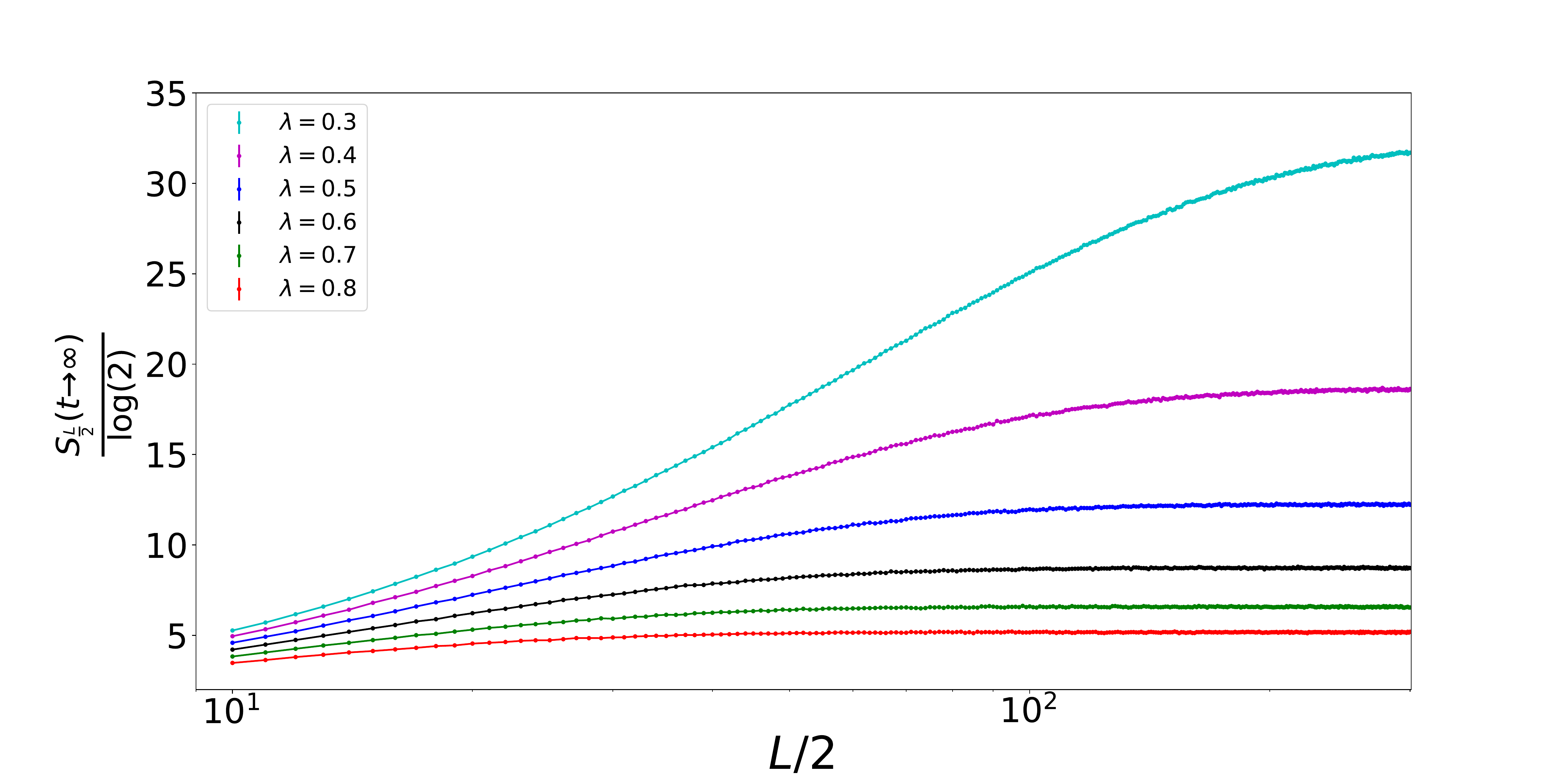}
\caption{$S_{\frac{L}{2}}$ plotted against system size at $t = 10^{11}$. Each
  point is an average over $5000$  random field realizations and the error
  shown is the standard error on the calculated mean. System sizes are taken
  from $L=20$ to $L=600$. Note, the logarithmic x-axis.
} \label{fig:ententropyvsL}
\end{figure}
	
In Fig. \ref{fig:ententropyvsL} we show results for the entanglement entropy versus $L/2$ at $t =
10^{11}$ as we vary the system size. We observe that as the system size is increased
the slope  of  $S_{\frac{L}{2}}(t \to \infty)$ is not constant. Instead, $S_{\frac{L}{2}}(t \to \infty)$
is indeed approaching a constant value as we increase system size. This means the
system is approaching an area law as the system size significantly exceeds the
localization length consistent with other studies~\cite{Abdul-Rahman2016,RahmanXY}. However, as is particularly
evident for $\lambda=0.3$, there can be an extended range of system sizes for which $S$ is linear in $\log(L)$.
		
\section{Localization Length} \label{Sec:Loc}
In this section we use the data from Fig.~ \ref{fig:ententropyvsL} to define a
quantity $\xi$ which is a measure of the localization length in the RFXX. We say the system is
completely localized when the entanglement entropy between our two subsystems
does not grow as we increase the system. When $L$ is small, unless disorder is
extremely large, we expect the entanglement entropy to grow sub-linearly in $L$
but it will still grow. So by adding one site to each sub-system, we grow the
lattice and determine the slope of $S_{\frac{L}{2}}$ with $L/2$. We can then define the
rate of growth, 
\begin{equation}
m(L/2) := S_{\frac{L}{2}}-S_{\frac{L}{2}-1}.
\end{equation}
In the localized regime we expect that, 
\begin{equation}
		\lim_{L \to \infty} m(L/2) = 0.
\end{equation}
The data however is not strictly increasing due to noise, so to improve the fitting
we use a Savitzky-–Golay filter to smooth the data and compute $m(L/2)$
with the smoothed version of the data. Defining a tolerance $\epsilon$, such
that $m(L/2)<\epsilon$ we can then define $\xi(\lambda)= \frac{L}{2}$ by the first $L$ for which 
this occurs. We choose $\epsilon$ to be reasonably small, since it indicates that the
function $m(L/2)$ is approaching the area law. 
\begin{figure}[!ht]
\centering
\includegraphics[width=\linewidth]{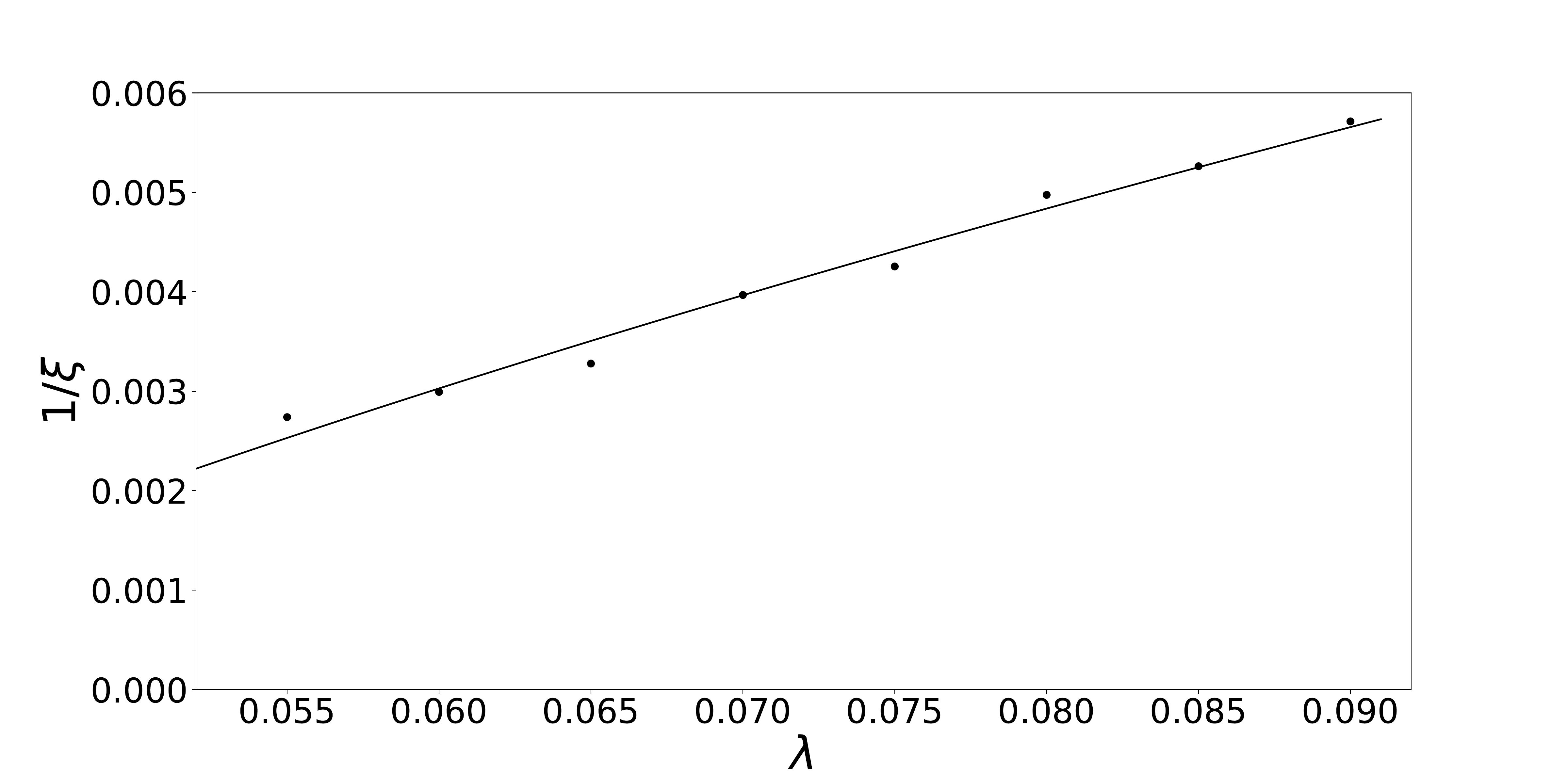}
\caption{ 
$\frac{1}{\xi}$ plotted against $\lambda$. The data from $S_{\frac{L}{2}}$ was
  smoothed out using a Savitzky-–Golay with a polynomial of degree two and a
  window of eleven, and a tolerance $\epsilon = 0.37$. Each value of
  $S_{\frac{L}{2}}$ was computed with over 20,000 realizations of the
  Hamiltonian. }
\label{fig:loclength}
\end{figure}
Our results are shown in Fig.~\ref{fig:loclength} clearly indicating a
diverging $\xi$ as $\lambda\to 0$. The fitted function takes the form
$a\sqrt{x}+b$ with standard deviations on the variables smaller than $3\times
10^{-3}$. The value of $b$ was found to be $b = -0.00866331$ and we expect this
value to approach zero as values closer to $\lambda = 0$ are probed. It is at present not
clear how reliable the above analysis is for a precise determination of the critical exponents,
but the results strongly suggest a diverging length scale as $\lambda\to 0$.
	
\section{Conclusion}
The presence of disorder in the RFXX has been shown to significantly alter the behaviour of the OTOCs. At a finite
disorder dependent $\xi_{OTOC}$ information propagation stops and the OTOCs are essentially zero beyond this length
scale. However, for $|x|<\xi_{OTOC}$ we find propagation at the maximal speed $v=J$ and confirm a power-law behaviour
for the early-time regime of $C(x,t)\sim t^{2x}$ with a position dependent exponent. An analysis of the behaviour
of $C(x,t)$ close to the wave-front shows a behaviour that is not consistent with recent predictions. The growth
of the entanglement starting from an un-entangled product state shows saturation at sufficiently large times.
We have not been able to isolate any specific temperature dependent effects and in the light of a temperature dependent
maximal bound on the Lyaponov exponents, $\lambda_L \le 2\pi k_BT/\hbar$, further studies would be of interest.

Finally, our results shed some light on the connection between thermalization and scrambling.
We observed {\it weak} scrambling in the localized phase ($\lambda\neq 0$) of the RFXX. From the results of Ref.~\cite{Gramsch2012}
it is known that relaxation in a closely related model is described by a  generalized Gibbs ensemble with
an {\it extensive} number of conserved quantities.
We also observe an absence of scrambling in the non-disordered ($\lambda=0$) case which requires an {\it intensive} number of
conserved quantities in the corresponding generalized Gibbs ensemble~\cite{Gluza2018}.  Hence, the absence of scrambling does not
imply the absence of a generalized form of thermalization and a sign of ``weak"
scrambling does not imply thermalization in the traditional sense.
	
\section{Acknowledgements}
This research was supported by NSERC and 
enabled in part by support provided by (SHARCNET) (www.sharcnet.ca) and Compute/Calcul Canada (www.computecanada.ca).

\onecolumngrid
\appendix
\section{Time Evolving Free Fermions}
\label{TimeEvolve}
In this appendix we review how to time evolve free fermions. A similar
treatment can be found in \cite{timeevolvefermions}. Starting from the
Hamiltonian, 
\begin{equation}
 \hat{H} = \sum_{i,j} M_{i,j}\hat{f}_i^{\dagger}\hat{f}_j,
\end{equation}
where we $M$ is a real $L\times L$ symmetric matrix and for generality we do not make any other assumption. This model represents a one dimensional system of quasi-free fermions hopping on a lattice. 
The fermionic operators  $\hat{f}_i^{\dagger}$ and $\hat{f}_i$ obey the anti-commutation relations, 
\begin{equation}
\{\hat f^\dagger_j,\hat f_k\}=\delta_{jk},\enspace \{\hat f^\dagger_j,\hat f^\dagger_k\}=\{\hat f_j,\hat f_k\}=0.
\end{equation}
Since $M$ is real symmetric we can always diagonalize it as $M=ADA^T $ where $AA^T = \mathbb{I}$ is real orthogonal transformation and D is a diagonal matrix with entries $D_{k,k} = \epsilon_k$ which are (real) energy eigenmodes. Defining new fermion operators,
\begin{equation}
\hat{d}_k = \sum_j {A}_{j,k} \hat{f}_j,
\end{equation}
\begin{equation}
\hat{d}_k^\dagger = \sum_j A_{j,k} \hat{f}_j^\dagger,
\end{equation}	
we can write the Hamiltonian as, 
\begin{equation}
\hat{H} = \sum_{k}\epsilon_{k}\hat{d}_{k}^\dagger\hat{d}_{k},
\end{equation}
	
The above operators can be referred to as reciprocal space  or normal modes operators.
These operators inherit fermionic anti-commutation relations due to the unitary property of $A$,
\begin{equation}
\{\hat{d}_l,\hat{d}_k^\dagger\} = \sum_{i,j} A_{i,l}{A}_{j,k}\{\hat{f}_i,\hat{f}_j^\dagger\} = \delta_{l,k}.
\end{equation}
Due to the definition of the annihilation operators it is easy to see that $|0\rangle_f = |0\rangle_d$. Thus all eigenstates can be constructed by applying creation operators $\hat{d}_k^\dagger$. These states are Gaussian, meaning they are completely described by their second moments.  Gaussian states can be completely described by the occupation matrix, $\Lambda_{i,j}^f=\langle \hat{f}_i^\dagger \hat{f}_j \rangle $ or in eigenmode space $\Lambda_{l,k}^d = \langle\hat{d}_l^\dagger \hat{d}_k \rangle$. All time evolved properties of this model can similarly be deduced by time evolving the occupation matrix. It is simple to time evolve the operators in eigenmode space, 
	 
\begin{equation}
\frac{d}{dt} (\hat{d_k} ) = i [\hat{H},\hat{d}_k],
\end{equation}
where, 
\begin{equation}
\hat{H} = \sum_k \epsilon_k \hat{d}_k^\dagger \hat{d}_k.
\end{equation}
Using, $\{\hat{d}_k,\hat{d}_l^\dagger\}=\delta_{l,k}$ and $\hat{d}_k^2 = 0$ one finds that, 
\begin{equation}
\hat{d}_k(t) = e^{-i\epsilon_kt}\hat{d}_k,
\end{equation}
similarly for the creation operators,
\begin{equation}
\hat{d}_k(t)^\dagger = e^{i\epsilon_kt}\hat{d}_k^\dagger,
\end{equation} 
this then implies,
\begin{equation}
\Lambda^d(t) = e^{iDt}\Lambda^d e^{-iDt}.
\end{equation}
Which means if we know, $\Lambda^d(0) = \Lambda^d$ we can compute $\Lambda^d(t)$ giving us all 
two point correlators taken at identical times.
Because we want to extract local statistics we need to transform back to the local fermion space. 
We see this is done by the following transformation,
\begin{equation} \label{occmatrix}
\Lambda^f(t) = A e^{iDt} \Lambda^d e^{-iDt}A^T,
\end{equation}
where, $\Lambda^d = A^T \Lambda^f A$. 
Now since we will also be interested in out of time correlations, 
it becomes important to consider two point correlations which are taken at different times.
For this we introduce the following notation, $\Lambda^f(t,t)$ 
where the left $t$ argument indicates that the creation operators $\hat{d}_k^\dagger$ are 
at a time $t$ and the right for the annihilation operators. 
Thus Eq. (\ref{occmatrix}) is $\Lambda^f(t) = \Lambda^f(t,t) $ and the out of time two point correlators are given by,
\begin{eqnarray}
\label{quenchdynamics}
\Lambda^{f}(t,t) &=& A e^{iDt} \Lambda^{d} e^{-iDt}  A^T, \\ 
\Lambda^{f}(t,0) &=& A e^{iDt} \Lambda^{d}  A^T, \\
\label{quenchdynamics3}
\Lambda^{f}(0,t) &=& A  \Lambda^{d} e^{-iDt}  A^T.
\end{eqnarray}
With Eqs.~(\ref{quenchdynamics}) to (\ref{quenchdynamics3})  we can calculate any two point correlator that might be expressed in the OTOC.
Next, it is important to see how the anti-commutation rule behaves as we consider creation and annihilation operators at different times.
In local space, consider the case where one operator in the Heisenberg picture is taken at $t=0$ and the other at $t=t$, 
\begin{equation}
\{ \hat{f}_m^\dagger(t), \hat{f}_n^\dagger\} 
= \sum_{k,l} {A}_{n,l} {A}_{m,k}e^{i\epsilon_k t}( \hat{d}_k^\dagger \hat{d}_l ^\dagger + \hat{d}_l^\dagger \hat{d}_k^\dagger ) = 0.
\end{equation}
Similarly $ \{ \hat{f}_m(t), \hat{f}_n\} =0 $ however the anti-commutation between
out of time creation and annihilation operators is non trivial, 
\begin{equation}
\label{anticom}
\{ \hat{f}_m^\dagger(t), \hat{f}_n\} 
= \sum_{k} {A}_{m,k} A_{n,k}e^{i \epsilon_k t} = a_{m,n}(t).
\end{equation}
At $t=0$ we see, $a_{m,n}(0) = \delta_{m,n}$ but time evolution removes this nice behaviour. We also see that, 
\begin{equation}
\bar{a}_{m,n} (t)  = 	\{ \hat{f}_m (t), \hat{f}_n^\dagger\} = \sum_{k} {A}_{m,k} A_{n,k}e^{-i \epsilon_k t}.
\end{equation}
With these tools in place it is convenient to write down the correlations exactly which will be featured in the OTOC. Consider two sites on the lattice labelled by $i$ and $j$ at $t=t$ and $t=0$ respectfully, then the time dependent correlations are taken from entries of 
Eqs.~(\ref{quenchdynamics}) to (\ref{quenchdynamics3}),  
\begin{eqnarray}
\label{corrs}
  \Lambda^{f}(t,t)_{i,i}&=&\langle \hat{f}_i^\dagger(t) \hat{f}_i(t) \rangle =  \sum_{k,l} e^{i(\epsilon_k-\epsilon_l) t} A_{i,k}A_{i,l}\langle \hat{d}_k^\dagger \hat{d}_l \rangle, \\
  \Lambda^{f}(t,0)_{i,j}&=&\langle \hat{f}_i^\dagger(t) \hat{f}_j \rangle =  \sum_{k,l}  e^{i\epsilon_k t}A_{i,k} A_{j,l} \langle \hat{d}_k^\dagger \hat{d}_l \rangle,   \\
  \Lambda^{f}(0,t)_{j,i}&=&\langle \hat{f}_j^\dagger \hat{f}_i(t) \rangle=   \sum_{k,l}  e^{-i\epsilon_l t}A_{j,k} A_{i,l} \langle \hat{d}_k^\dagger \hat{d}_l \rangle, \\
\label{corrs4}
  \Lambda^{f}(0,0)_{j,j}&=&\langle \hat{f}_j^\dagger \hat{f}_j \rangle =  \sum_{k,l} A_{j,k}A_{j,l}\langle \hat{d}_k^\dagger \hat{d}_l \rangle.
\end{eqnarray}
With this we have all the ingredients we require to compute an OTOC. 
In the case of a thermal state or an eigenstate the expressions in Eqs. (\ref{corrs}) to (\ref{corrs4}) 
are greatly simplified since the occupation matrix in eigenmode space is diagonal. We consider a Gibbs state of the form, 
\begin{equation} 
\rho = \frac{e^{-\beta \hat{H}}}{Z}.
\end{equation} 
	 
For thermal states we label the correlations with an additional $\beta$. The correlations in eigenmode space are well known with different sites decoupled and the occupation numbers following a Fermi-Dirac statistic with zero chemical potential,
\begin{equation}
\label{thermOcc}
\Lambda_{k,l}^{d,\beta} = \langle \hat{d}_k^\dagger \hat{d}_l \rangle_\beta =  \left\{
\begin{array}{cl}
\frac{1}{1+e^{\beta \epsilon_k}} & k=l,  \\
0 &\text{otherwise}.
\end{array}\right.
\end{equation}
In the next appendix section we describe how to use these expressions to compute the OTOC between two $S^z$ operators on different sites.

\section{Out of time order correlations} \label{OTOCcalcs}
	
The OTOC we compute in section \ref{Sec:OTOCs} relies on the computation of the Eq. (\ref{fOTOC}), or rewriting it here, 
\begin{equation}
F(t) = \langle \hat{\sigma}_i^z(t) \hat{\sigma}_j^z \hat{\sigma}_i^z(t) \hat{\sigma}_j^z  \rangle.
\end{equation}
Where we have dropped the $x=|i-j|$ term in favour of expressing it as only a function of time. Evaluating this expression is the same as evaluating Eq. (\ref{otoc}). For the following it is easy to represent, $\hat{n}_i(t) = \hat{f}_i^\dagger(t)\hat{f}_i(t)$. Substituting the Jordan-Wigner transformation definition, 
\begin{equation}
F(t) = 16\langle (\hat{n}_{i}(t)-\frac{1}{2} )(\hat{n}_{j}-\frac{1}{2} )(\hat{n}_{i}(t)-\frac{1}{2} )(\hat{n}_{j}-\frac{1}{2} ) \rangle.
\end{equation}
Expanding this and simplifying this using $\hat{n}_i(t)^2 = n_i(t)$ 
and the anti-commutation rules shown in Eq. (\ref{anticom}) we can write,
\begin{eqnarray}
\label{fsimp}
F(t) =16 \langle \hat{n}_{i}(t) \hat{n}_{j} \hat{n}_{i}(t) \hat{n}_{j} 
-\frac{1}{2}(\hat{n}_{i}(t) \hat{n}_{j} \hat{n}_{i}(t) +   \hat{n}_{j} \hat{n}_{i}(t) \hat{n}_{j} ) 
+ \frac{1}{4}(\hat{n}_{j} \hat{n}_{i}(t) - \hat{n}_{i}(t) \hat{n}_{j} ) + \frac{1}{16} \rangle. 
\end{eqnarray}
	 
Using Eq. (\ref{fsimp}) we can now use the definitions of our initial conditions on $\Lambda^{d}$ to derive exact expressions for the OTOCs.
\subsection{Product States}
We consider our initial state as one constructed from the vacuum state such that, 
\begin{equation}
|\Psi\rangle = \prod_{j \in \mathbb{S}}\hat{f}_j^\dagger|0\rangle
\end{equation}
Where the cardinality of the set $\mathbb{S}$ represents the conserved 
number of fermions on the lattice, $\langle \hat{N} \rangle = \sum_j  \langle\hat{f}_j^\dagger \hat{f}_j\rangle = |\mathbb{S}|$.  
This gives us an initial local occupation matrix of the form, 
\begin{equation}
\Lambda_{i,j}^{f}(0) = \langle \hat{f}_i^\dagger \hat{f}_j \rangle =  \left\{
\begin{array}{cl}
1 & i=j \ \wedge i\in \mathbb{S}  \\
0 &\text{otherwise}.
\end{array}\right.
\end{equation}
	
First consider the case that $\hat{\sigma}_j^z$ is selected such that $j\in \mathbb{S}$.  Then using  $\hat{f}_j^\dagger |\psi \rangle = 0$ and Eq. (\ref{anticom}) we get, 
\begin{equation}
\label{prodF}
F(t) = 8 |a_{i,j}(t)|^2 \langle \hat{n}_{i}(t)\rangle -8|a_{i,j}(t)|^2 + 1. 
\end{equation}
Similarly if we assume $j\notin \mathbb{S}$ such that $\hat{f}_j |\psi \rangle = 0$ then we recover, 
\begin{equation}\label{ProdF2}
F(t) =  1 -  8|a_{i,j}(t)|^2 \langle \hat{n}_i(t) \rangle. 
\end{equation}
Eqs. (\ref{prodF}) and (\ref{ProdF2}) reveal that the fundamental behaviour of the OTOC relies on $|a_{i,j}(t)|^2$ and $\langle \hat{n}_{i}(t)\rangle$. The product state OTOC will have two effects coming together, equilibration of  $\langle \hat{n}_{i}(t)\rangle$ and the out of time anti-commutation relation $|a_{i,j}(t)|^2$. This extra equilibration is expected to contribute to extra structure not present in the thermal case.

\subsection{Thermal States}
The thermal OTOC is computed similarly to the product state, 
but we exploit its simple structure in eigenmode space as seen in Eqs. (\ref{thermOcc}). 
Here we exploit the fact that $\hat{f}_i^2 = \hat{f}_i^{\dagger2} = 0$ and use Wicks theorem for thermal states \cite{GaudinWick}. 
This gives us the following form,
\begin{eqnarray}
\label{thermF}
F(t) = 16|a_{i,j}(t)|^2\left( \langle \hat{f}_i^\dagger \hat{f}_i\rangle_\beta  \langle \hat{f}_j^\dagger \hat{f}_j\rangle_\beta  
          -\frac{1}{2}\left(\langle \hat{f}_i^\dagger \hat{f}_i\rangle_\beta+  \langle \hat{f}_j^\dagger \hat{f}_j\rangle_\beta \right) 
          +  \bar{a}_{i,j}(t)\langle \hat{f}_i^\dagger(t)\hat{f}_j\rangle_\beta-\langle \hat{f}_i^\dagger(t)\hat{f}_j\rangle_\beta \langle \hat{f}_j^\dagger\hat{f}_i(t)\rangle_\beta \right) +1.
\end{eqnarray}
Where we have used the fact that same time two point correlators are stationary, $\langle \hat{f}_i^\dagger(t)\hat{f}_i(t)\rangle_\beta = \langle \hat{f}_i^\dagger\hat{f}_i\rangle_\beta$. 	Eq. (\ref{thermF}) is quite a bit more complicated than Eq. (\ref{prodF}) but the defining behaviour is still reliant on $|a_{i,j}(t)|^2$ while  the quantity  $\langle \hat{n}_{i}(t)\rangle$ is now time independent. Instead we see out of time correlations in the form of $\langle \hat{f}_j^\dagger\hat{f}_i(t)\rangle_\beta $ for example play a role.

\twocolumngrid
\bibliography{references}

\begin{thebibliography}{57}
\expandafter\ifx\csname natexlab\endcsname\relax\def\natexlab#1{#1}\fi
\expandafter\ifx\csname bibnamefont\endcsname\relax
  \def\bibnamefont#1{#1}\fi
\expandafter\ifx\csname bibfnamefont\endcsname\relax
  \def\bibfnamefont#1{#1}\fi
\expandafter\ifx\csname citenamefont\endcsname\relax
  \def\citenamefont#1{#1}\fi
\expandafter\ifx\csname url\endcsname\relax
  \def\url#1{\texttt{#1}}\fi
\expandafter\ifx\csname urlprefix\endcsname\relax\def\urlprefix{URL }\fi
\providecommand{\bibinfo}[2]{#2}
\providecommand{\eprint}[2][]{\url{#2}}

\bibitem[{\citenamefont{Maldacena et~al.}(2016)\citenamefont{Maldacena,
  Shenker, and Stanford}}]{Maldacena2016a}
\bibinfo{author}{\bibfnamefont{J.}~\bibnamefont{Maldacena}},
  \bibinfo{author}{\bibfnamefont{S.~H.} \bibnamefont{Shenker}},
  \bibnamefont{and} \bibinfo{author}{\bibfnamefont{D.}~\bibnamefont{Stanford}},
  \bibinfo{journal}{Journal of High Energy Physics}
  \textbf{\bibinfo{volume}{2016}}, \bibinfo{pages}{106} (\bibinfo{year}{2016}).

\bibitem[{\citenamefont{Larkin and Ovchinnikov}(1969)}]{Larkin1969}
\bibinfo{author}{\bibfnamefont{A.~I.} \bibnamefont{Larkin}} \bibnamefont{and}
  \bibinfo{author}{\bibfnamefont{Y.~N.} \bibnamefont{Ovchinnikov}},
  \bibinfo{journal}{Zh. Eksp. Teor. Fiz.} \textbf{\bibinfo{volume}{55}},
  \bibinfo{pages}{2262} (\bibinfo{year}{1969}), \bibinfo{note}{[JETP 28, 1200
  (1969)]}.

\bibitem[{\citenamefont{Swingle and Chowdhury}(2017)}]{Swingle2017}
\bibinfo{author}{\bibfnamefont{B.}~\bibnamefont{Swingle}} \bibnamefont{and}
  \bibinfo{author}{\bibfnamefont{D.}~\bibnamefont{Chowdhury}},
  \bibinfo{journal}{Phys. Rev. B} \textbf{\bibinfo{volume}{95}},
  \bibinfo{pages}{060201} (\bibinfo{year}{2017}).

\bibitem[{\citenamefont{Sekino and Susskind}(2008)}]{Sekino2008}
\bibinfo{author}{\bibfnamefont{Y.}~\bibnamefont{Sekino}} \bibnamefont{and}
  \bibinfo{author}{\bibfnamefont{L.}~\bibnamefont{Susskind}},
  \bibinfo{journal}{Journal of High Energy Physics}
  \textbf{\bibinfo{volume}{2008}}, \bibinfo{pages}{065} (\bibinfo{year}{2008}).

\bibitem[{\citenamefont{Sachdev and Ye}(1993)}]{Sachdev1993}
\bibinfo{author}{\bibfnamefont{S.}~\bibnamefont{Sachdev}} \bibnamefont{and}
  \bibinfo{author}{\bibfnamefont{J.}~\bibnamefont{Ye}}, \bibinfo{journal}{Phys.
  Rev. Lett.} \textbf{\bibinfo{volume}{70}}, \bibinfo{pages}{3339}
  (\bibinfo{year}{1993}).

\bibitem[{\citenamefont{Sachdev}(2015)}]{Sachdev2015}
\bibinfo{author}{\bibfnamefont{S.}~\bibnamefont{Sachdev}},
  \bibinfo{journal}{Phys. Rev. X} \textbf{\bibinfo{volume}{5}},
  \bibinfo{pages}{041025} (\bibinfo{year}{2015}).

\bibitem[{\citenamefont{Roberts et~al.}(2015)\citenamefont{Roberts, Stanford,
  and Susskind}}]{Roberts2015a}
\bibinfo{author}{\bibfnamefont{D.~A.} \bibnamefont{Roberts}},
  \bibinfo{author}{\bibfnamefont{D.}~\bibnamefont{Stanford}}, \bibnamefont{and}
  \bibinfo{author}{\bibfnamefont{L.}~\bibnamefont{Susskind}},
  \bibinfo{journal}{Journal of High Energy Physics}
  \textbf{\bibinfo{volume}{2015}}, \bibinfo{pages}{51} (\bibinfo{year}{2015}).

\bibitem[{\citenamefont{Fu and Sachdev}(2016)}]{Fu2016}
\bibinfo{author}{\bibfnamefont{W.}~\bibnamefont{Fu}} \bibnamefont{and}
  \bibinfo{author}{\bibfnamefont{S.}~\bibnamefont{Sachdev}},
  \bibinfo{journal}{Phys. Rev. B} \textbf{\bibinfo{volume}{94}},
  \bibinfo{pages}{035135} (\bibinfo{year}{2016}).

\bibitem[{\citenamefont{Maldacena and Stanford}(2016)}]{Maldacena2016b}
\bibinfo{author}{\bibfnamefont{J.}~\bibnamefont{Maldacena}} \bibnamefont{and}
  \bibinfo{author}{\bibfnamefont{D.}~\bibnamefont{Stanford}},
  \bibinfo{journal}{Phys. Rev. D} \textbf{\bibinfo{volume}{94}},
  \bibinfo{pages}{106002} (\bibinfo{year}{2016}).

\bibitem[{\citenamefont{D\'ora and Moessner}(2017)}]{Dora2017}
\bibinfo{author}{\bibfnamefont{B.}~\bibnamefont{D\'ora}} \bibnamefont{and}
  \bibinfo{author}{\bibfnamefont{R.}~\bibnamefont{Moessner}},
  \bibinfo{journal}{Phys. Rev. Lett.} \textbf{\bibinfo{volume}{119}},
  \bibinfo{pages}{026802} (\bibinfo{year}{2017}).

\bibitem[{\citenamefont{Huang et~al.}(2017)\citenamefont{Huang, Zhang, and
  Chen}}]{Huang2017}
\bibinfo{author}{\bibfnamefont{Y.}~\bibnamefont{Huang}},
  \bibinfo{author}{\bibfnamefont{Y.-L.} \bibnamefont{Zhang}}, \bibnamefont{and}
  \bibinfo{author}{\bibfnamefont{X.}~\bibnamefont{Chen}},
  \bibinfo{journal}{Annalen der Physik} \textbf{\bibinfo{volume}{529}},
  \bibinfo{pages}{1600318} (\bibinfo{year}{2017}).

\bibitem[{\citenamefont{Chen et~al.}(2017)\citenamefont{Chen, Zhou, Huse, and
  Fradkin}}]{Chen2017}
\bibinfo{author}{\bibfnamefont{X.}~\bibnamefont{Chen}},
  \bibinfo{author}{\bibfnamefont{T.}~\bibnamefont{Zhou}},
  \bibinfo{author}{\bibfnamefont{D.~A.} \bibnamefont{Huse}}, \bibnamefont{and}
  \bibinfo{author}{\bibfnamefont{E.}~\bibnamefont{Fradkin}},
  \bibinfo{journal}{Annalen der Physik} \textbf{\bibinfo{volume}{529}},
  \bibinfo{pages}{1600332} (\bibinfo{year}{2017}).

\bibitem[{\citenamefont{Slagle et~al.}(2017)\citenamefont{Slagle, Bi, You, and
  Xu}}]{Slagle2017}
\bibinfo{author}{\bibfnamefont{K.}~\bibnamefont{Slagle}},
  \bibinfo{author}{\bibfnamefont{Z.}~\bibnamefont{Bi}},
  \bibinfo{author}{\bibfnamefont{Y.-Z.} \bibnamefont{You}}, \bibnamefont{and}
  \bibinfo{author}{\bibfnamefont{C.}~\bibnamefont{Xu}}, \bibinfo{journal}{Phys.
  Rev. B} \textbf{\bibinfo{volume}{95}}, \bibinfo{pages}{165136}
  (\bibinfo{year}{2017}).

\bibitem[{\citenamefont{Fan et~al.}(2017)\citenamefont{Fan, Zhang, Shen, and
  Zhai}}]{Fan2017}
\bibinfo{author}{\bibfnamefont{R.}~\bibnamefont{Fan}},
  \bibinfo{author}{\bibfnamefont{P.}~\bibnamefont{Zhang}},
  \bibinfo{author}{\bibfnamefont{H.}~\bibnamefont{Shen}}, \bibnamefont{and}
  \bibinfo{author}{\bibfnamefont{H.}~\bibnamefont{Zhai}},
  \bibinfo{journal}{Science Bulletin} \textbf{\bibinfo{volume}{62}},
  \bibinfo{pages}{707 } (\bibinfo{year}{2017}), ISSN \bibinfo{issn}{2095-9273}.

\bibitem[{\citenamefont{Deng et~al.}(2017)\citenamefont{Deng, Li, Pixley, Wu,
  and Das~Sarma}}]{Deng2017}
\bibinfo{author}{\bibfnamefont{D.-L.} \bibnamefont{Deng}},
  \bibinfo{author}{\bibfnamefont{X.}~\bibnamefont{Li}},
  \bibinfo{author}{\bibfnamefont{J.~H.} \bibnamefont{Pixley}},
  \bibinfo{author}{\bibfnamefont{Y.-L.} \bibnamefont{Wu}}, \bibnamefont{and}
  \bibinfo{author}{\bibfnamefont{S.}~\bibnamefont{Das~Sarma}},
  \bibinfo{journal}{Phys. Rev. B} \textbf{\bibinfo{volume}{95}},
  \bibinfo{pages}{024202} (\bibinfo{year}{2017}).

\bibitem[{\citenamefont{Nandkishore and Huse}(2015)}]{Nandkishore2015}
\bibinfo{author}{\bibfnamefont{R.}~\bibnamefont{Nandkishore}} \bibnamefont{and}
  \bibinfo{author}{\bibfnamefont{D.~A.} \bibnamefont{Huse}},
  \bibinfo{journal}{Annual Review of Condensed Matter Physics}
  \textbf{\bibinfo{volume}{6}}, \bibinfo{pages}{15} (\bibinfo{year}{2015}).

\bibitem[{\citenamefont{Alet and Laflorencie}(2018)}]{Alet2018}
\bibinfo{author}{\bibfnamefont{F.}~\bibnamefont{Alet}} \bibnamefont{and}
  \bibinfo{author}{\bibfnamefont{N.}~\bibnamefont{Laflorencie}},
  \bibinfo{journal}{Comptes Rendus Physique}  (\bibinfo{year}{2018}), ISSN
  \bibinfo{issn}{1631-0705}.

\bibitem[{\citenamefont{Luitz and Bar~Lev}(2017)}]{Luitz2017}
\bibinfo{author}{\bibfnamefont{D.~J.} \bibnamefont{Luitz}} \bibnamefont{and}
  \bibinfo{author}{\bibfnamefont{Y.}~\bibnamefont{Bar~Lev}},
  \bibinfo{journal}{Phys. Rev. B} \textbf{\bibinfo{volume}{96}},
  \bibinfo{pages}{020406} (\bibinfo{year}{2017}).

\bibitem[{\citenamefont{Xu and Swingle}(2018{\natexlab{a}})}]{Xu2018a}
\bibinfo{author}{\bibfnamefont{S.}~\bibnamefont{Xu}} \bibnamefont{and}
  \bibinfo{author}{\bibfnamefont{B.}~\bibnamefont{Swingle}},
  \bibinfo{journal}{arXiv.org}  (\bibinfo{year}{2018}{\natexlab{a}}),
  \eprint{1805.05376v1}.

\bibitem[{\citenamefont{Xu and Swingle}(2018{\natexlab{b}})}]{Xu2018b}
\bibinfo{author}{\bibfnamefont{S.}~\bibnamefont{Xu}} \bibnamefont{and}
  \bibinfo{author}{\bibfnamefont{B.}~\bibnamefont{Swingle}},
  \bibinfo{journal}{arXiv.org}  (\bibinfo{year}{2018}{\natexlab{b}}),
  \eprint{1802.00801}.

\bibitem[{\citenamefont{Sahu et~al.}(2018)\citenamefont{Sahu, Xu, and
  Swingle}}]{Sahu2018}
\bibinfo{author}{\bibfnamefont{S.}~\bibnamefont{Sahu}},
  \bibinfo{author}{\bibfnamefont{S.}~\bibnamefont{Xu}}, \bibnamefont{and}
  \bibinfo{author}{\bibfnamefont{B.}~\bibnamefont{Swingle}},
  \bibinfo{journal}{arXiv} \textbf{\bibinfo{volume}{cond-mat.str-el}}
  (\bibinfo{year}{2018}).

\bibitem[{\citenamefont{Shenker and Stanford}(2014)}]{Shenker2014a}
\bibinfo{author}{\bibfnamefont{S.~H.} \bibnamefont{Shenker}} \bibnamefont{and}
  \bibinfo{author}{\bibfnamefont{D.}~\bibnamefont{Stanford}},
  \bibinfo{journal}{Journal of High Energy Physics}
  \textbf{\bibinfo{volume}{2014}}, \bibinfo{pages}{67} (\bibinfo{year}{2014}).

\bibitem[{\citenamefont{Gu et~al.}(2017)\citenamefont{Gu, Qi, and
  Stanford}}]{Gu2017}
\bibinfo{author}{\bibfnamefont{Y.}~\bibnamefont{Gu}},
  \bibinfo{author}{\bibfnamefont{X.-L.} \bibnamefont{Qi}}, \bibnamefont{and}
  \bibinfo{author}{\bibfnamefont{D.}~\bibnamefont{Stanford}},
  \bibinfo{journal}{Journal of High Energy Physics}
  \textbf{\bibinfo{volume}{2017}}, \bibinfo{pages}{125} (\bibinfo{year}{2017}),
  ISSN \bibinfo{issn}{1029-8479}.

\bibitem[{\citenamefont{Patel et~al.}(2017)\citenamefont{Patel, Chowdhury,
  Sachdev, and Swingle}}]{Patel2017}
\bibinfo{author}{\bibfnamefont{A.~A.} \bibnamefont{Patel}},
  \bibinfo{author}{\bibfnamefont{D.}~\bibnamefont{Chowdhury}},
  \bibinfo{author}{\bibfnamefont{S.}~\bibnamefont{Sachdev}}, \bibnamefont{and}
  \bibinfo{author}{\bibfnamefont{B.}~\bibnamefont{Swingle}},
  \bibinfo{journal}{Phys. Rev. X} \textbf{\bibinfo{volume}{7}},
  \bibinfo{pages}{031047} (\bibinfo{year}{2017}).

\bibitem[{\citenamefont{Chowdhury and Swingle}(2017)}]{Chowdhury2017}
\bibinfo{author}{\bibfnamefont{D.}~\bibnamefont{Chowdhury}} \bibnamefont{and}
  \bibinfo{author}{\bibfnamefont{B.}~\bibnamefont{Swingle}},
  \bibinfo{journal}{Phys. Rev. D} \textbf{\bibinfo{volume}{96}},
  \bibinfo{pages}{065005} (\bibinfo{year}{2017}).

\bibitem[{\citenamefont{Nahum et~al.}(2018)\citenamefont{Nahum, Vijay, and
  Haah}}]{Nahum2018}
\bibinfo{author}{\bibfnamefont{A.}~\bibnamefont{Nahum}},
  \bibinfo{author}{\bibfnamefont{S.}~\bibnamefont{Vijay}}, \bibnamefont{and}
  \bibinfo{author}{\bibfnamefont{J.}~\bibnamefont{Haah}},
  \bibinfo{journal}{Physical Review X} \textbf{\bibinfo{volume}{8}},
  \bibinfo{pages}{021014} (\bibinfo{year}{2018}).

\bibitem[{\citenamefont{Khemani et~al.}(2018)\citenamefont{Khemani, Vishwanath,
  and Huse}}]{Khemani2018}
\bibinfo{author}{\bibfnamefont{V.}~\bibnamefont{Khemani}},
  \bibinfo{author}{\bibfnamefont{A.}~\bibnamefont{Vishwanath}},
  \bibnamefont{and} \bibinfo{author}{\bibfnamefont{D.}~\bibnamefont{Huse}},
  \bibinfo{journal}{Physical Review X} \textbf{\bibinfo{volume}{8}},
  \bibinfo{pages}{031057} (\bibinfo{year}{2018}).

\bibitem[{\citenamefont{Rakovszky et~al.}(2018)\citenamefont{Rakovszky,
  Pollmann, and von Keyserlingk}}]{Rakovsky2018}
\bibinfo{author}{\bibfnamefont{T.}~\bibnamefont{Rakovszky}},
  \bibinfo{author}{\bibfnamefont{F.}~\bibnamefont{Pollmann}}, \bibnamefont{and}
  \bibinfo{author}{\bibfnamefont{C.~W.} \bibnamefont{von Keyserlingk}},
  \bibinfo{journal}{Phys. Rev. X} \textbf{\bibinfo{volume}{8}},
  \bibinfo{pages}{031058} (\bibinfo{year}{2018}).

\bibitem[{\citenamefont{Liu and Suh}(2014)}]{Liu2014}
\bibinfo{author}{\bibfnamefont{H.}~\bibnamefont{Liu}} \bibnamefont{and}
  \bibinfo{author}{\bibfnamefont{S.~J.} \bibnamefont{Suh}},
  \bibinfo{journal}{Phys. Rev. Lett.} \textbf{\bibinfo{volume}{112}},
  \bibinfo{pages}{011601} (\bibinfo{year}{2014}).

\bibitem[{\citenamefont{Hosur et~al.}(2016)\citenamefont{Hosur, Qi, Roberts,
  and Yoshida}}]{Hosur2016}
\bibinfo{author}{\bibfnamefont{P.}~\bibnamefont{Hosur}},
  \bibinfo{author}{\bibfnamefont{X.-L.} \bibnamefont{Qi}},
  \bibinfo{author}{\bibfnamefont{D.~A.} \bibnamefont{Roberts}},
  \bibnamefont{and} \bibinfo{author}{\bibfnamefont{B.}~\bibnamefont{Yoshida}},
  \bibinfo{journal}{Journal of High Energy Physics}
  \textbf{\bibinfo{volume}{2016}}, \bibinfo{pages}{4} (\bibinfo{year}{2016}),
  ISSN \bibinfo{issn}{1029-8479}.

\bibitem[{\citenamefont{Yunger~Halpern}(2017)}]{Halpern2017a}
\bibinfo{author}{\bibfnamefont{N.}~\bibnamefont{Yunger~Halpern}},
  \bibinfo{journal}{Phys. Rev. A} \textbf{\bibinfo{volume}{95}},
  \bibinfo{pages}{012120} (\bibinfo{year}{2017}).

\bibitem[{\citenamefont{Yunger~Halpern
  et~al.}(2018)\citenamefont{Yunger~Halpern, Swingle, and
  Dressel}}]{Halpern2017b}
\bibinfo{author}{\bibfnamefont{N.}~\bibnamefont{Yunger~Halpern}},
  \bibinfo{author}{\bibfnamefont{B.}~\bibnamefont{Swingle}}, \bibnamefont{and}
  \bibinfo{author}{\bibfnamefont{J.}~\bibnamefont{Dressel}},
  \bibinfo{journal}{Phys. Rev. A} \textbf{\bibinfo{volume}{97}},
  \bibinfo{pages}{042105} (\bibinfo{year}{2018}).

\bibitem[{\citenamefont{Alonso et~al.}(2018)\citenamefont{Alonso,
  Yunger~Halpern, and Dressel}}]{Alonso2018}
\bibinfo{author}{\bibfnamefont{J.~R.~G.} \bibnamefont{Alonso}},
  \bibinfo{author}{\bibfnamefont{N.}~\bibnamefont{Yunger~Halpern}},
  \bibnamefont{and} \bibinfo{author}{\bibfnamefont{J.}~\bibnamefont{Dressel}},
  \bibinfo{journal}{arXiv.org}  (\bibinfo{year}{2018}), \eprint{1806.09637}.

\bibitem[{\citenamefont{Bardarson et~al.}(2012)\citenamefont{Bardarson,
  Pollmann, and Moore}}]{BardarsonEnt}
\bibinfo{author}{\bibfnamefont{J.~H.} \bibnamefont{Bardarson}},
  \bibinfo{author}{\bibfnamefont{F.}~\bibnamefont{Pollmann}}, \bibnamefont{and}
  \bibinfo{author}{\bibfnamefont{J.~E.} \bibnamefont{Moore}},
  \bibinfo{journal}{Phys. Rev. Lett.} \textbf{\bibinfo{volume}{109}},
  \bibinfo{pages}{017202} (\bibinfo{year}{2012}).

\bibitem[{\citenamefont{Serbyn et~al.}(2013)\citenamefont{Serbyn,
  Papi\ifmmode~\acute{c}\else \'{c}\fi{}, and Abanin}}]{Serbyn2013}
\bibinfo{author}{\bibfnamefont{M.}~\bibnamefont{Serbyn}},
  \bibinfo{author}{\bibfnamefont{Z.}~\bibnamefont{Papi\ifmmode~\acute{c}\else
  \'{c}\fi{}}}, \bibnamefont{and} \bibinfo{author}{\bibfnamefont{D.~A.}
  \bibnamefont{Abanin}}, \bibinfo{journal}{Phys. Rev. Lett.}
  \textbf{\bibinfo{volume}{110}}, \bibinfo{pages}{260601}
  (\bibinfo{year}{2013}).

\bibitem[{\citenamefont{Abdul-Rahman et~al.}(2016)\citenamefont{Abdul-Rahman,
  Nachtergaele, Sims, and Stolz}}]{Abdul-Rahman2016}
\bibinfo{author}{\bibfnamefont{H.}~\bibnamefont{Abdul-Rahman}},
  \bibinfo{author}{\bibfnamefont{B.}~\bibnamefont{Nachtergaele}},
  \bibinfo{author}{\bibfnamefont{R.}~\bibnamefont{Sims}}, \bibnamefont{and}
  \bibinfo{author}{\bibfnamefont{G.}~\bibnamefont{Stolz}},
  \bibinfo{journal}{Letters in Mathematical Physics}
  \textbf{\bibinfo{volume}{106}}, \bibinfo{pages}{649} (\bibinfo{year}{2016}),
  ISSN \bibinfo{issn}{1573-0530}.

\bibitem[{\citenamefont{von Keyserlingk et~al.}(2018)\citenamefont{von
  Keyserlingk, Rakovszky, Pollmann, and Sondhi}}]{VonKeyserlingk2018}
\bibinfo{author}{\bibfnamefont{C.~W.} \bibnamefont{von Keyserlingk}},
  \bibinfo{author}{\bibfnamefont{T.}~\bibnamefont{Rakovszky}},
  \bibinfo{author}{\bibfnamefont{F.}~\bibnamefont{Pollmann}}, \bibnamefont{and}
  \bibinfo{author}{\bibfnamefont{S.~L.} \bibnamefont{Sondhi}},
  \bibinfo{journal}{Phys. Rev. X} \textbf{\bibinfo{volume}{8}},
  \bibinfo{pages}{021013} (\bibinfo{year}{2018}).

\bibitem[{\citenamefont{Bohrdt et~al.}(2017)\citenamefont{Bohrdt, Mendl,
  Endres, and Knap}}]{Bohrdt2017}
\bibinfo{author}{\bibfnamefont{A.}~\bibnamefont{Bohrdt}},
  \bibinfo{author}{\bibfnamefont{C.~B.} \bibnamefont{Mendl}},
  \bibinfo{author}{\bibfnamefont{M.}~\bibnamefont{Endres}}, \bibnamefont{and}
  \bibinfo{author}{\bibfnamefont{M.}~\bibnamefont{Knap}}, \bibinfo{journal}{New
  Journal of Physics} \textbf{\bibinfo{volume}{19}}, \bibinfo{pages}{063001}
  (\bibinfo{year}{2017}).

\bibitem[{\citenamefont{Lewis-Swan et~al.}(2018)}]{LewisSwan2018}
\bibinfo{author}{\bibfnamefont{R.~J.} \bibnamefont{Lewis-Swan}}
  \bibnamefont{et~al.}, \bibinfo{journal}{arXiv.org}  (\bibinfo{year}{2018}),
  \eprint{1808.07134}.

\bibitem[{\citenamefont{Chaitanya~Murthy}(2018)}]{Murthy2018}
\bibinfo{author}{\bibfnamefont{M.~S.} \bibnamefont{Chaitanya~Murthy}},
  \bibinfo{journal}{arXiv.org}  (\bibinfo{year}{2018}), \eprint{1809.03681}.

\bibitem[{\citenamefont{Marek~Gluza}(2018)}]{Gluza2018}
\bibinfo{author}{\bibfnamefont{T.~F.} \bibnamefont{Marek~Gluza},
  \bibfnamefont{Jens~Eisert}}, \bibinfo{journal}{arXiv.org}
  (\bibinfo{year}{2018}), \eprint{1809.08268}.

\bibitem[{\citenamefont{Lin and Motrunich}(2018{\natexlab{a}})}]{LinOTOCising}
\bibinfo{author}{\bibfnamefont{C.-J.} \bibnamefont{Lin}} \bibnamefont{and}
  \bibinfo{author}{\bibfnamefont{O.~I.} \bibnamefont{Motrunich}},
  \bibinfo{journal}{Phys. Rev. B} \textbf{\bibinfo{volume}{97}},
  \bibinfo{pages}{144304} (\bibinfo{year}{2018}{\natexlab{a}}).

\bibitem[{\citenamefont{Byju et~al.}(2018)\citenamefont{Byju, Lochan, and
  Shankaranarayanan}}]{Byju2018}
\bibinfo{author}{\bibfnamefont{S.}~\bibnamefont{Byju}},
  \bibinfo{author}{\bibfnamefont{K.}~\bibnamefont{Lochan}}, \bibnamefont{and}
  \bibinfo{author}{\bibfnamefont{S.}~\bibnamefont{Shankaranarayanan}},
  \bibinfo{journal}{arXiv.org}  (\bibinfo{year}{2018}), \eprint{1808.07742}.

\bibitem[{\citenamefont{Lin and Motrunich}(2018{\natexlab{b}})}]{Lin2018}
\bibinfo{author}{\bibfnamefont{C.-J.} \bibnamefont{Lin}} \bibnamefont{and}
  \bibinfo{author}{\bibfnamefont{O.~I.} \bibnamefont{Motrunich}},
  \bibinfo{journal}{arXiv.org}  (\bibinfo{year}{2018}{\natexlab{b}}),
  \eprint{1807.08826v1}.

\bibitem[{\citenamefont{Hamza et~al.}(2012)\citenamefont{Hamza, Sims, and
  Stolz}}]{Hamza2012}
\bibinfo{author}{\bibfnamefont{E.}~\bibnamefont{Hamza}},
  \bibinfo{author}{\bibfnamefont{R.}~\bibnamefont{Sims}}, \bibnamefont{and}
  \bibinfo{author}{\bibfnamefont{G.}~\bibnamefont{Stolz}},
  \bibinfo{journal}{Communications in Mathematical Physics}
  \textbf{\bibinfo{volume}{315}}, \bibinfo{pages}{215} (\bibinfo{year}{2012}),
  ISSN \bibinfo{issn}{1432-0916}.

\bibitem[{\citenamefont{Coleman}(2015)}]{Coleman}
\bibinfo{author}{\bibfnamefont{P.}~\bibnamefont{Coleman}},
  \emph{\bibinfo{title}{Introduction to Many-Body Physics}}
  (\bibinfo{publisher}{Cambridge University Press}, \bibinfo{year}{2015}).

\bibitem[{\citenamefont{Riddell and M\"uller}(2018)}]{RiddellGETH}
\bibinfo{author}{\bibfnamefont{J.}~\bibnamefont{Riddell}} \bibnamefont{and}
  \bibinfo{author}{\bibfnamefont{M.~P.} \bibnamefont{M\"uller}},
  \bibinfo{journal}{Phys. Rev. B} \textbf{\bibinfo{volume}{97}},
  \bibinfo{pages}{035129} (\bibinfo{year}{2018}).

\bibitem[{\citenamefont{Lai and Yang}(2015)}]{Lai2015}
\bibinfo{author}{\bibfnamefont{H.-H.} \bibnamefont{Lai}} \bibnamefont{and}
  \bibinfo{author}{\bibfnamefont{K.}~\bibnamefont{Yang}},
  \bibinfo{journal}{Phys. Rev. B} \textbf{\bibinfo{volume}{91}},
  \bibinfo{pages}{081110} (\bibinfo{year}{2015}).

\bibitem[{\citenamefont{Abdul-Rahman et~al.}(2017)\citenamefont{Abdul-Rahman,
  Nachtergaele, Sims, and Stolz}}]{RahmanXY}
\bibinfo{author}{\bibfnamefont{H.}~\bibnamefont{Abdul-Rahman}},
  \bibinfo{author}{\bibfnamefont{B.}~\bibnamefont{Nachtergaele}},
  \bibinfo{author}{\bibfnamefont{R.}~\bibnamefont{Sims}}, \bibnamefont{and}
  \bibinfo{author}{\bibfnamefont{G.}~\bibnamefont{Stolz}},
  \bibinfo{journal}{Annalen der Physik} \textbf{\bibinfo{volume}{529}},
  \bibinfo{pages}{1600280} (\bibinfo{year}{2017}).

\bibitem[{\citenamefont{Stolz}(2011)}]{StolzALintro}
\bibinfo{author}{\bibfnamefont{G.}~\bibnamefont{Stolz}},
  \bibinfo{journal}{arXiv preprint arXiv:1104.2317}  (\bibinfo{year}{2011}).

\bibitem[{\citenamefont{Miller}(1972)}]{wmillersymmetry}
\bibinfo{author}{\bibfnamefont{W.}~\bibnamefont{Miller}},
  \emph{\bibinfo{title}{Symmetry Groups and Their Applications}}, Computer
  Science and Applied Mathematics (\bibinfo{publisher}{Academic Press},
  \bibinfo{year}{1972}), ISBN \bibinfo{isbn}{9780124974609}.

\bibitem[{\citenamefont{Peschel and Eisler}(2009)}]{PeschelEE}
\bibinfo{author}{\bibfnamefont{I.}~\bibnamefont{Peschel}} \bibnamefont{and}
  \bibinfo{author}{\bibfnamefont{V.}~\bibnamefont{Eisler}},
  \bibinfo{journal}{Journal of Physics A: Mathematical and Theoretical}
  \textbf{\bibinfo{volume}{42}}, \bibinfo{pages}{504003}
  (\bibinfo{year}{2009}).

\bibitem[{\citenamefont{Pouranvari et~al.}(2015)\citenamefont{Pouranvari,
  Zhang, and Yang}}]{Pouranvari2015}
\bibinfo{author}{\bibfnamefont{M.}~\bibnamefont{Pouranvari}},
  \bibinfo{author}{\bibfnamefont{Y.}~\bibnamefont{Zhang}}, \bibnamefont{and}
  \bibinfo{author}{\bibfnamefont{K.}~\bibnamefont{Yang}},
  \bibinfo{journal}{Advances in Condensed Matter Physics}
  \textbf{\bibinfo{volume}{2015}} (\bibinfo{year}{2015}).

\bibitem[{\citenamefont{Latorre and Riera}(2009)}]{LatorreEE}
\bibinfo{author}{\bibfnamefont{J.~I.} \bibnamefont{Latorre}} \bibnamefont{and}
  \bibinfo{author}{\bibfnamefont{A.}~\bibnamefont{Riera}},
  \bibinfo{journal}{Journal of Physics A: Mathematical and Theoretical}
  \textbf{\bibinfo{volume}{42}}, \bibinfo{pages}{504002}
  (\bibinfo{year}{2009}).

\bibitem[{\citenamefont{Perarnau-Llobet
  et~al.}(2016)\citenamefont{Perarnau-Llobet, Riera, Gallego, Wilming, and
  Eisert}}]{timeevolvefermions}
\bibinfo{author}{\bibfnamefont{M.}~\bibnamefont{Perarnau-Llobet}},
  \bibinfo{author}{\bibfnamefont{A.}~\bibnamefont{Riera}},
  \bibinfo{author}{\bibfnamefont{R.}~\bibnamefont{Gallego}},
  \bibinfo{author}{\bibfnamefont{H.}~\bibnamefont{Wilming}}, \bibnamefont{and}
  \bibinfo{author}{\bibfnamefont{J.}~\bibnamefont{Eisert}},
  \bibinfo{journal}{New Journal of Physics} \textbf{\bibinfo{volume}{18}},
  \bibinfo{pages}{123035} (\bibinfo{year}{2016}).

\bibitem[{\citenamefont{Gramsch and Rigol}(2012)}]{Gramsch2012}
\bibinfo{author}{\bibfnamefont{C.}~\bibnamefont{Gramsch}} \bibnamefont{and}
  \bibinfo{author}{\bibfnamefont{M.}~\bibnamefont{Rigol}},
  \bibinfo{journal}{Phys. Rev. A} \textbf{\bibinfo{volume}{86}},
  \bibinfo{pages}{053615} (\bibinfo{year}{2012}).

\bibitem[{\citenamefont{Gaudin}(1960)}]{GaudinWick}
\bibinfo{author}{\bibfnamefont{M.}~\bibnamefont{Gaudin}},
  \bibinfo{journal}{Nuclear Physics} \textbf{\bibinfo{volume}{15}},
  \bibinfo{pages}{89 } (\bibinfo{year}{1960}).

\end{thebibliography}
\end{document}